\documentclass[aps,prb,twocolumn,showpacs,preprintnumbers,amsmath,amssymb,superscriptaddress]{revtex4-2}
\usepackage{epsfig,amsopn}
\usepackage{graphicx}
\usepackage{epstopdf}
\usepackage{sidecap}
\usepackage{color}
\usepackage{amsmath,amssymb}
\usepackage{amsfonts}
\usepackage{textcomp}
\usepackage{enumerate}
\usepackage{soul}
\usepackage{float}
\usepackage{scalerel}
\usepackage{braket}
\usepackage{tikz}
\usepackage{natbib}
\usepackage{booktabs}
\usepackage[ colorlinks = true,
linkcolor = red,
urlcolor  = red,
citecolor = blue,
anchorcolor = red]{hyperref}

\newcommand{\beq}{\begin{equation}}
  \newcommand{\eeq}{\end{equation}}
\newcommand{\bqa}{\begin{eqnarray}}
  \newcommand{\eqa}{\end{eqnarray}}

\newcommand{\moire}{moir$\mathrm{\acute{e}}\;$}

\begin{document}
	\title{Charge and spin orders of the generalized Wigner crystal phases at fractional fillings in twisted MoTe$_2$: unrestricted Hartree-Fock and variational Monte Carlo studies }
	  \author{Arindam Pramanik}
        \email{apramanik@agh.edu.pl}
        \affiliation{Academic Centre for Materials and Nanotechnology,
          AGH University of Krakow, Al. Mickiewicza 30, 30-059 Krakow, Poland }
          \author{Andrzej Biborski}
          \affiliation{Academic Centre for Materials and Nanotechnology,
          	AGH University of Krakow, Al. Mickiewicza 30, 30-059 Krakow, Poland }
	\author{Michał Zegrodnik}
    \email{michal.zegrodnik@agh.edu.pl}
	\affiliation{Academic Centre for Materials and Nanotechnology,
		AGH University of Krakow, Al. Mickiewicza 30, 30-059 Krakow, Poland }
	%
	%
	%
	\begin{abstract}
		Recent experiments have revealed interaction-driven insulating states at various fractional fillings of twisted bilayer MoTe$_2$. Many of these insulating states are believed to be generalized Wigner crystals (GWC)  condensed in the background of triangular moir$\acute{e}$ lattice. In this work, we focus on revealing the charge and spin order of the insulating states at 4/3 and 3/2 hole fillings over a range of twist angles and applied displacement fields using unrestricted Hartree-Fock (HF) and variational Monte Carlo (VMC) methods. We find that, at 4/3 filling for low twist angles the charge ordered ground state is valley-polarized (VP), characterized by weak charge delocalization approaching the classical limit. Application of the displacement field induces layer-polarization and suppresses the charge fluctuations even further. In contrast, at 3/2 filling, the VP phase exhibits signatures of a two-component GWC phase, which is particularly robust at finite displacement field. Additionally, by increasing the twist angle one pushes the system into an intervalley-coherent (IVC) phase, which is found to be fluctuation-dominated. Our VMC-based analysis shows that the correlation energy is negligible in the VP phase, which is consistent with its classical nature. However, the IVC phase shows significant gain in correlation energy, which is dominated by the nonlocal correlations captured by the Jastrow factors, while the local Gutzwiller factors play a minor role.

	\end{abstract}

	\maketitle
        
 \section{Introduction}
 
In 1934, Eugene Wigner predicted that electrons moving in an inert neutralizing background would spontaneously form a crystalline phase at sufficiently low density where the Coulomb interaction energy dominates over the kinetic energy \cite{Wigner_PR_1934}. The first experimental realization of WC phase came more than 40 years after its prediction, in 1979 in the electrons on the liquid He surface \cite{Grimes_PRL_1979}. Later in GaAs/GaAlAs 2DEG, the WC was realized by  quenching the kinetic energy in the presence of an external magnetic field \cite{Andrei_PRL_1988}. Very recently, a direct observation of the WC phase has been achieved in Bernal-stacked bilayer graphene under external magnetic field \cite{Tsui_Nature_2024}. Without the external magnetic field, the WC phase has been observed in monolayer MoSe$_2$ \cite{Smolenski_Nature_2021} and a heterostructure of MoSe$_2$, where two monolayers are separated by hBN \cite{Zhou_Nature_2021}. Additionally, the advent of \moire materials has allowed to realize such a phase even without the need of an external magnetic field. In the \moire flat bands, carriers experience a large Coulomb interaction, which can lead to the formation of the WC phase at certain carrier densities, as experimentally shown in bilayer WSe$_2$-WS$_2$ at fractional fillings $n=1/3$, 1/2 and 2/3 \cite{Xu_Nature_2020, HLi_Nature_2021, XHuang_NatPhys_2021}, as well as in twisted WS$_2$ \cite{Li_Science_2024}. From a theoretical perspective, multiple studies have confirmed or predicted the WC phase in various \moire systems \cite{Padhi_PRB_2021, Pan_PRB_2020, Matty_NatCommun_2022, Duran_PRB_2023, Ung_PRB_2023, Tsang_PRR_2023, ZSong_PRB_2023, YZhou_PRL_2024, Erkensten_PRB_2024, Kumar_npj_2025, Biborski_PRB_2025, Biborski_Polonica_2026, Mishra_arxiv_2026, YZhou_arxiv_2025, Saha_PRB_2025, Peng_PRB_2026, Chen_SciBull_2026, Wang_PRB_2026, Tsang_PRR_2023}.
It should be noted that the mentioned realizations of the WC phase are not exactly the phase Wigner predicted. Namely, in contrast to the original concept, the \moire materials possess an underlying lattice, which constrains the real-space motion of the electrons and consequently influences the order of the resultant Wigner crystal. In the literature, such lattice-modified WC phases are referred to as generalized Wigner Crystals (GWCs) \cite{Hubbard_PRB_1978, Noda_PRL_2002}.

Similar to GWC, the charge density wave (CDW) phenomenon also exhibits spontaneous charge modulation in real space. However, the CDW phase can even appear in the weak-coupling regime as a result of the Fermi surface instability due to the nesting effect \cite{Gruner_1994, Gruner_RevModPhys_1988}, while the WC/GWC state is due to substantial interactions leading to charge localization. Additionally, the charge density in GWC shows strong real-space inhomogeneity, which is commensurate with the undrelying lattice and appears at specific band fillings. On the other hand, in CDW a much smoother modulation around the average value is seen and both commensurate and incommensurate periodicities can be realized within a wider range of band fillings also coexisting with a metallic state. 

Due to a significant Coulomb repulsion and large degree of tunability, transition metal dichalcogenide (TMD) \moire systems can be considered as potential hosts of generalized Wigner-crystal phases with distinct charge and spin configurations. 
Apart from the structures previously studied in the context of possible GWC realization, experiments on twisted bilayer WSe$_2$ (t-WSe$_2$) and bilayer MoTe$_2$ (t-MoTe$_2$) have revealed 
numerous insulating states beyond fractional Chern insulator and anomalous quantum Hall phase, which are induced by tuning the experimentally accessible parameters such as the carrier density, displacement field, and magnetic field \cite{Knuppel_NatCommun_2025, Xu_arxiv_2026, Wolff_arxiv_2026}. Importantly, those insulating states appear at specific band fillings which together with significant strength of inter-site Coulomb repulsion points to a possible interaction-driven charge localization and formation of the GWC.

In this work, we investigate the charge and spin orders in t-MoTe$_2$ at hole fillings $n=4/3$ and $3/2$, where insulating states have been experimentally observed \cite{Xu_arxiv_2026} and are mostly unexplored by theoretical calculations. These states have been excluded as topologically nontrivial insulators such as fractional Chern insulators, making them plausible candidates for generalized Wigner-crystal phases. We study the 2-band tight-binding model over a broad range of twist angles and displacement fields using Hartree-Fock (HF) and variational Monte Carlo (VMC) methods. The phase diagrams and associated charge and spin textures are established through extensive HF analysis. We discuss the origin of the subsequent symmetry-broken states and their principal features. In general, the obtained results can be divided into two types of states: the valley polarized (VP) and intervalley coherent (IVC) state. The former appears at lower twist angles and corresponds to a dominant role of electron-electron interactions, and a classical character, while the latter is realized in the case of larger angles, and is characterized by an increased role of kinetic energy and dominant fluctuations indicating its quantum nature. By comparing the VMC and HF results, we discuss the role of electron-electron correlations and show that while in the case of VP state the two methods give virtually the same results, after the transition to IVC state the differences become clearly apparent. Additionally, by selectively optimizing different parts of the variational wavefunction within VMC, we try to disentangle the contributions of different variational parameters to the correlation energy of the system.

The remainder of the paper is organized as follows. Section II, introduces the model Hamiltonian of the system and discusses the HF and VMC methodologies. In section III, the phase diagrams along with the details of the charge and spin textures of the phases are presented. Section IV discusses the VMC results and the effects of correlation beyond HF. Finally, section V summarizes the main findings of the paper.

\section{Model and Method}

\subsection{Model Hamiltonian}
\begin{figure}[t]
    \vspace{6pt}
    \centering
    \includegraphics[width=0.49\textwidth]{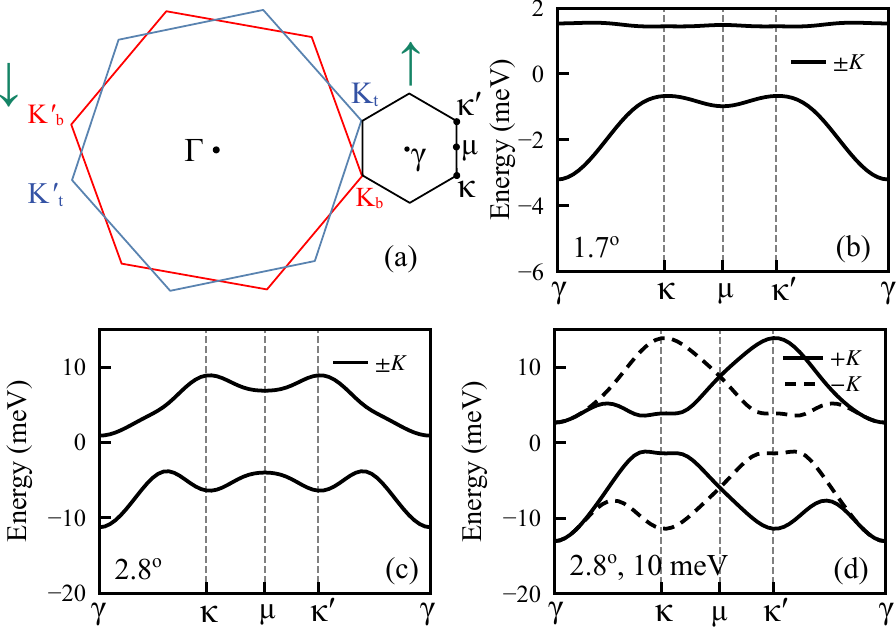}
    \caption{
    (a) Schematic representation of BZs. The blue and red hexagons denote the BZs of the top and bottom monolayers, respectively. The black hexagon denotes the mBZ of t-MoTe$_2$ system. The uparrow and downarrow in the $K$ and $K'(-K)$ valley represent the spin-valley locking property. (b),(c) The bandstructures of $K(+K)$ and $K' (-K)$ valleys for $\theta=1.7^\circ$, and $2.8^\circ$, respectively at $\Delta=0$. (d) The bandstructure of $\theta=2.8^\circ$ at $\Delta=10$ meV.
    }
    \label{fig_bands}
\end{figure}

The strong Ising spin-orbit coupling in monolayer MoTe$_2$ leads to spin-valley locking. As a result, the low-energy valence states near the $K$ and $K'$ valleys of the monolayer Brillouin zone are primarily spin-$\uparrow$ and spin-$\downarrow$ states, respectively. The twisted bilayer system inherits this spin-valley locking property \cite{Wu_Lovorn_PRL_2019}. Consequently, in t-MoTe$_2$, the low-energy \moire states originating from the $K$ and $K'$ valleys of the monolayer Brillouin zone can be identified with the spin-$\uparrow$ and spin-$\downarrow$ sectors, respectively. Therefore, within the low-energy subspace of t-MoTe$_2$, spin and valley degrees of freedom are effectively locked to each other. Hence, throughout this work, we use the terms spin and valley interchangeably. The schematic in Fig.~\ref{fig_bands}(a) shows the Brillouin zones of the top and bottom monolayers, denoted by blue and red hexagons, respectively, and the \moire Brillouin zone (mBZ) is denoted by the black hexagon. In this work, we retain the two highest \moire valence bands from each valley as active bands and model the system using the two-band tight-binding model introduced in \cite{Cheng_PNAS_2024}. Within such an approach the two bands result from two Wannier orbitals residing in MX and XM stacking configurations of the original moir\'e lattice, where $M$ and $X$ correspond to metal (Mo) and chalcogen (Te) ions, respectively. This leads to two interpenetrating triangular sublattices denoted here by $A$ and $B$, which together form a honeycomb structure. The resulting single-particle Hamiltonian is provided below.

\begin{equation}
\begin{split}
H_{0} &=\sum_{n,m,\alpha,\sigma} \sum_{r=\{1,3\}}t_r\, \hat{c}^{\dagger}_{n\alpha\sigma} \hat{c}_{m\bar{\alpha} \sigma}\\
&\quad + \sum_{n,m,\alpha,\beta,\sigma} \sum_{r=\{2,4,5,6\}} |t_r| e^{i\tilde{\sigma}\nu_{n\alpha,m\beta}\phi}\, \hat{c}^{\dagger}_{n\alpha \sigma} \hat{c}_{m\beta \sigma}\;,
\end{split}
\label{eq_H0}
\end{equation}
where $\hat{c}^{\dagger}_{n\alpha\sigma}$ ($\hat{c}_{n\alpha\sigma}$) creates (annihilates) an electron with spin $\sigma$ in the sublattice $\alpha$ in unit cell $n$ of the honeycomb lattice. The spin index $\sigma$ represents the two spin components $\uparrow$ and $\downarrow$. Additionally, $\tilde{\sigma}=1$ ($\tilde{\sigma}=-1$) for $\sigma=\uparrow$ ($\sigma=\downarrow$) and $\bar{\alpha}=B$ ($\bar{\alpha}=A$) for $\alpha=A$ ($\alpha=B$). The hopping terms are taken up to the $6^{th}$ nearest neighbor with amplitudes $t_r$ corresponding to the $r$-th nearest neighbor among which only $t_1$ and $t_3$ are purely real. The remaining ones are complex, with the complex phase depending both on spin and hopping direction. Namely, $\nu_{n\alpha,m\beta}$ is +1 or -1 depending on the relative orientation of $(n,\alpha)$ and $(m,\beta)$, i.e, the two sites involved in the hopping process. For the sign convention of $\nu_{n\alpha,m\beta}$ refer to the supplementary material \cite{supple}. The noninteracting bands of t-MoTe$_2$ obtained by diagonalizing Hamiltonian of Eq.~\eqref{eq_H0} and are shown in Figs.~\ref{fig_bands}(b) and  (c) for selected values of the twist angles ($\theta$) $1.7^\circ$ and $2.8^\circ$, respectively. The bands originating from the two valleys $K$ and $K'$ exhibit degeneracy due to the time-reversal symmetry and the emergent inversion symmetry of the two-band tight-binding model. As can be seen, the bandwidth of the two bands increases with the twist angle.

It should be noted that the Wannier orbitals of the \moire system centered on the $A$ and $B$ sublattices are predominantly localized in two different monolayers \cite{Cheng_PNAS_2024}. Therefore, within the present model, the layer and sublattice degrees of freedom are effectively locked, and we use the terms layer and sublattice interchangeably throughout this work. Due to this correspondence, the effect of an out-of-plane displacement field is incorporated through an on-site energy difference between the $A$ and $B$ sublattices.
\begin{equation}
H_D=-\frac{\Delta}{2} \sum_{n, \alpha\in A, \sigma}  \hat{c}^{\dagger}_{n\alpha\sigma} \hat{c}_{n\alpha \sigma}+\frac{\Delta}{2} \sum_{n,\alpha\in B, \sigma}  \hat{c}^{\dagger}_{n\alpha \sigma} \hat{c}_{n\alpha \sigma}\;,
\label{eq_HD}
\end{equation}
where $\Delta=\frac{Dd}{\epsilon}$, $d$ is the distance between layers, $D$ is the displacement field, and $\epsilon$ is the dielectric constant. The effect of the displacement field on the band structure is shown in Fig.~\ref{fig_bands} (d) for a selected value of the twist angle $\theta=2.8^\circ$. As one can see the displacement field destroys the emergent inversion symmetry of the two-band tight-binding model and lifts the valley degeneracy. It also increases the bandwidth of the individual bands.

To take into account interaction within the effective two-band model, the Coulomb term is projected on the top two noninteracting valence bands of each valley, which are considered as active bands in the study. The projection generates many terms, including onsite Coulomb, intersite Coulomb, ferromagnetic exchange, and a three-site correlated hopping \cite{Cheng_PNAS_2024}. The intersite Coulomb and the ferromagnetic exchange are considered up to the $3^{rd}$ and $2^{nd}$ nearest neighbors, respectively. Taking into account all these terms, the effective interaction Hamiltonian takes the following form.
\begin{equation}
\begin{split}
H_{int} &=U\sum_{i} \hat{n}_{i\uparrow} \, \hat{n}_{i\downarrow}\\
&\quad +\sum_{i,j}^{3^{rd}NN}\sum_{\sigma,\sigma'} V_{ij}\,\hat{n}_{i\sigma}\, \hat{n}_{j\sigma'}\\
&\quad + \sum_{i,j}^{2^{nd}NN}   \left[\sum_{\sigma,\sigma'}\left(-\frac{1}{2}J_{ij}\,\hat{n}_{i\sigma}\, \hat{n}_{j\sigma'}\right)-2J_{ij}\hat{S}^z_i\hat{S}^z_j\right]\\
&\quad + \sum_{\ll i,j\gg, i'}\sum_{\sigma}T^{i'}_{ij}\hat{n}_{i'}\hat{c}^{\dagger}_{i\sigma} \hat{c}_{j\sigma},
\end{split}
\label{eq_Hint}
\end{equation}
where, $i,j$ denotes the combined unit cell $(n/m)$ and sublattice indices $(\alpha/\beta)$, while $U$ and $V$ represent the onsite and intersite density-density interaction, respectively. Here, $J_{ij}$ is the direct ferromagnetic spin exchange and the $z$-component of the spin operator is $\hat{S}^z_i=\frac{1}{2}\left[\hat{n}_{i\uparrow}-\hat{n}_{i,\downarrow} \right]$. The last term corresponds to correlated hopping with amplitude  $T^{i'}_{ij}$, where the hopping strength between two $2^{nd}$ nearest neighbor sites $i$ and $j$ depends on the particle density at the site $i'$, which is the common nearest neighbor of $i$ and $j$. The values of all the parameters mentioned in Eqs.~\eqref{eq_H0}, \eqref{eq_HD}, and ~\eqref{eq_Hint} for various $\theta$ values can be found in the supplementary material of Ref. \cite{Cheng_PNAS_2024}. 

We note that, although the filling fractions are referred to in terms of holes throughout this work, all calculations are performed in the electron picture.

\subsection{Unrestricted Hartree-Fock mean-field theory} 
     
We have used Hartree-Fock mean-field theory to study the total Hamiltonian,  $H=H_0+H_D+H_{int}$ given by Eqs.~\eqref{eq_H0}, ~\eqref{eq_HD}, and ~\eqref{eq_Hint}. In this method, the two-body (quartic) interaction terms are reduced to one-body (quadratic) terms by considering fluctuations up to the first order \cite{BruusFlensberg_2004}. Using this approximation we write $\hat{c}^\dagger_{i\sigma_1}\hat{c}_{j\sigma_2} \hat{c}^\dagger_{k\sigma_3}\hat{c}_{l\sigma_4}\approx \left\langle \hat{c}^\dagger_{i\sigma_1}\hat{c}_{j\sigma_2} \right\rangle \hat{c}^\dagger_{k\sigma_3}\hat{c}_{l\sigma_4} -  \left\langle \hat{c}^\dagger_{i\sigma_1}\hat{c}_{l\sigma_4} \right\rangle \hat{c}^\dagger_{k\sigma_3}\hat{c}_{j\sigma_2}$. With these simplifications, Eq.~\eqref{eq_Hint} takes the following mean-field form 
\begin{widetext}
\begin{equation}
\begin{split}
H^{MF}_{int} &=U\sum_{i} \left[\left\langle \hat{c}^\dagger_{i\uparrow}\hat{c}_{i\uparrow} \right\rangle \hat{c}^\dagger_{i\downarrow}\hat{c}_{i\downarrow} -  \left\langle \hat{c}^\dagger_{i\uparrow}\hat{c}_{i\downarrow} \right\rangle \hat{c}^\dagger_{i\downarrow}\hat{c}_{i\uparrow}\right]
+\sum_{i,j}^{3^{rd}NN}\sum_{\sigma,\sigma'} V_{ij}\,\left[\left\langle \hat{c}^\dagger_{i\sigma}\hat{c}_{i\sigma} \right\rangle \hat{c}^\dagger_{j\sigma'}\hat{c}_{j\sigma'} -  \left\langle \hat{c}^\dagger_{i\sigma}\hat{c}_{j\sigma'} \right\rangle \hat{c}^\dagger_{j\sigma'}\hat{c}_{i\sigma}\right]\\
&\quad - \sum_{i,j}^{2^{nd}NN}\sum_{\sigma} J_{ij}\,\left[\left\langle \hat{c}^\dagger_{i\sigma}\hat{c}_{i\sigma} \right\rangle \hat{c}^\dagger_{j\sigma}\hat{c}_{j\sigma} -  \left\langle \hat{c}^\dagger_{i\sigma}\hat{c}_{j\sigma} \right\rangle \hat{c}^\dagger_{j\sigma}\hat{c}_{i\sigma}\right] + \sum_{\ll i,j\gg, i'}\sum_{\sigma,\sigma'}T^{i'}_{ij}\left[\left\langle \hat{c}^\dagger_{i'\sigma'}\hat{c}_{i'\sigma'} \right\rangle \hat{c}^\dagger_{i\sigma}\hat{c}_{j\sigma} -  \left\langle \hat{c}^\dagger_{i'\sigma'}\hat{c}_{j\sigma} \right\rangle \hat{c}^\dagger_{i\sigma}\hat{c}_{i'\sigma'}\right]
\end{split}
\end{equation}	
\end{widetext}

The total mean-field Hamiltonian, $H^{MF}=H_0+H_D+H^{MF}_{int}$, is solved self-consistently to determine the ground state of the system. We have performed the self-consistent calculation in an unrestricted manner, i.e., without imposing any symmetry-related constraints on the mean fields, allowing for spontaneous symmetry breaking without assuming any predefined supercell structure. The unrestricted HF program of the mVMC package \cite{Misawa_CPC_2019} has been used for this purpose.

\subsection{Variational Monte Carlo}

In addition to using the HF mean-field method, we have employed the variational Monte Carlo method to determine the approximate ground state of the model Hamiltonian and to study the effects of the correlations beyond the mean-field approximation. The VMC method is one of the available tools for treating correlation physics beyond mean-field theory, where a suitable many-body variational wavefunction with many variational parameters is optimized to represent the ground state of the system \cite{BeccaSorella_2017, Toulouse_2016, Foulkes_RevModPhys_2001, Ceperley_PRB_1977, Yokoyama_Japan_1987, Biborski_PRB_2020}. The variational ansatz is given by 
\begin{equation}
|\Psi_T\rangle = \hat{\mathcal{P}}\hat{\mathcal{L}}|\Psi_0\rangle,
\end{equation}
where $|\Psi_0\rangle$ is the pair-product or Pfaffian wavefunction \cite{TaharaImada_2008, Misawa_CPC_2019}, which has the following form 
\begin{equation}
|\Psi_0\rangle = \left( \sum_{i,j,\sigma,\sigma'} f_{i\sigma,j\sigma'} \hat{c}^{\dagger}_{i\sigma} \hat{c}^{\dagger}_{j\sigma'}\right)^{N_e/2}|0\rangle,
\end{equation}
where $(i,j)$ are the combined unit cell and sublattice indices, $(\sigma,\sigma')$ are spin indices as introduced earlier. $N_e$ is the total number of electrons, which should be even, and $f_{i\sigma,j\sigma'}$'s are the variational parameters to be determined. The projector $\hat{\mathcal{P}}$ is used to describe correlation beyond the mean-field approximation, which contains factors like Gutzwiller and Jastrow. The projector $\hat{\mathcal{L}}$ denotes quantum-number projections like total-spin and momentum. The method of VMC is established on the variational principle, which states that 

\begin{equation}
    E_G \leq E_T=\frac{\langle \Psi_T|H|\Psi_T\rangle}{\langle \Psi_T|\Psi_T\rangle},
\end{equation}
where $E_G$ is the energy of the ground state. Any expectation value including the energy is computed by expanding $|\Psi_T\rangle$ using the basis \{$|x\rangle$\}, which is given by 

\begin{equation}
    |x\rangle= \prod_{i=1}^{N_e} \hat{c}^\dagger_{r_i\sigma_i}|0\rangle,
\end{equation}
where $r_i$ and $\sigma_i$ are the position and the spin of the $i^{th}$ electron. The configurations are sampled according to the probability density function $\rho(x) \propto |\langle x|\Psi_T\rangle|^2$ and $E_T$ is estimated as 

\begin{equation}
    E_T \approx \frac{1}{M} \sum_{l=1}^M \rho(x_l) \frac{\langle \Psi_T|H|x_l\rangle}{\langle \Psi_T|x_l\rangle},
\end{equation}
i.e., as an average over the $M$ samples. The expectation value of any operator $A$ is estimated similarly. 

For this work, in the projector $\hat{\mathcal{P}}$, we have considered only the Gutzwiller ($\hat{\mathcal{P_G}}$) and the Jastrow ($\hat{\mathcal{P_J}}$) factors, i.e. $\hat{\mathcal{P}}=\hat{\mathcal{P_G}} \hat{\mathcal{P_J}}$. These factors have the following forms.

\begin{equation}
\begin{aligned}
\hat{\mathcal{P_G}} &= e^{\sum_i g_i \, \hat{n}_{i\uparrow} \,\hat{n}_{i\downarrow}} \\
\hat{\mathcal{P_J}} &= e^{\frac{1}{2}{\sum_{i\neq j} v_{ij}(\hat{n}_i-1)(\hat{n}_j-1)}}
\end{aligned}
\label{eq:xy}
\end{equation}
where the value of the variational parameter $g_i$ controls the local correlations through the double occupancy at the site $i$ and the variational parameters $ v_{ij} $ introduce long-range correlation.  To obtain the minimum energy state, we have not imposed any symmetry-related constraints on the variational parameters $f_{i\sigma,j\sigma'}, g_i$, and $v_{ij}$. In this work, we have not considered the spin number projection. We also ignore the possibility of the backflow correlation. 

The VMC calculations have been performed using the mVMC package \cite{Misawa_CPC_2019}, which utilizes the stochastic reconfiguration method to optimize the variational parameters while minimizing the trial ground state energy $E_T$. 

We used a $6\times6$ lattice system for both the unrestricted HF and VMC calculations. Although this size of cluster may appear relatively small, every unit cell carries two orbital and two spin degrees of freedom, which results in a substantially large Hilbert space. Moreover, the VMC simulations are performed without imposing any symmetry constraints on the variational parameters; any larger systems than considered here are computationally very expensive. Also this choice of system size is sufficiently commensurate to accommodate a broad range of symmetry-breaking patterns, including $2\times1$, $2\times2$, $2\times3$, $\sqrt{3}\times\sqrt{3}$, and $3\times3$ reconstructions. Moreover, all the calculations correspond to zero temperature.

\section{Hartree-Fock Phase diagram}
We study various possible reconstructions of the charge and spin densities of the considered system in an unbiased manner using the unrestricted HF method. In our study, we primarily obtain two types of symmetry-broken phases, which we analyze in detail in the following. In the first case, there is a net imbalance in the spin-$\uparrow$ and spin-$\downarrow$ population, which results in an out-of-plane magnetization of the ground state leading to breaking of the time-reversal symmetry. Since the spin and valley degrees of freedom are locked, this situation corresponds to a valley polarized (VP) state. In the second case, the spin-$\uparrow$ and spin-$\downarrow$ sectors are in a coherent superposition, or equivalently, the valleys $K$ and $K'$ mix with each other, resulting in an intervalley coherent (IVC) phase with broken valley $U(1)$ symmetry of the model.

In the following sections, we present the phase diagrams and details of the charge and spin order of the respective phases in two selected fractional fillings ($n=4/3$ and $n=3/2$). The minimum energy state at a given $(\theta, \Delta)$ has been confirmed through multiple simulations, each starting with random initial mean fields.

\begin{figure*}[t]
    \centering
    \includegraphics[width=0.8\textwidth]{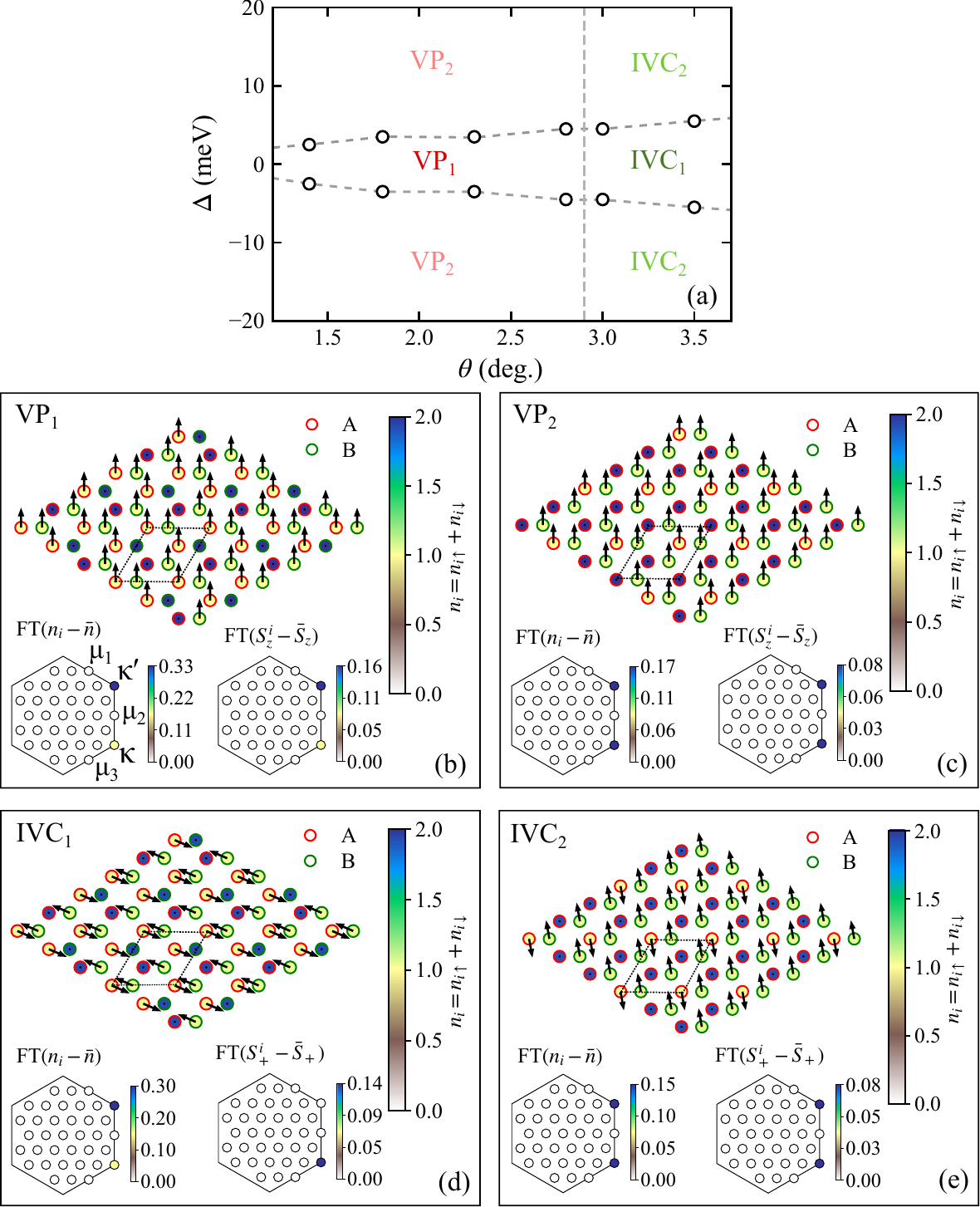}
    \caption{
    Phase diagram of hole-filling 4/3 case. (a) The phase of the ground state as a function of $\theta$ and $\Delta$. The open black circles denote the phase boundary determined from the HF calculations as a function of the displacement field at selected angles. The phases VP$_1$, VP$_2$, IVC$_1$, and IVC$_2$ are characterized in (b), (c), (d), and (e), respectively, for $\Delta > 0$. (b) The 6$\times$6 lattice system is shown by red ($A$ sublattice) and green ($B$ sublattice) circles. The colorbar shows the charge density. The arrows show spin. And the dashed lines depict the reconstructed unitcell. At the bottom, the left inset and the right inset plot the amplitudes of the FT of mean-subtracted charge and spin density, respectively. The details of (c), (d), and (e) are same as (b). Note that the arrows represent the $z$-component of spin in (b) \& (c) and the in-plane component of spin in (d) \& (e).
    }
    \label{fig_PD_-1.33}
\end{figure*}

\subsection{Hole-filling 4/3: phase diagram, charge, and spin order of the phases}
The phase diagram of the ground state as a function of $\theta$ and $\Delta$ at hole-filling 4/3 is presented in Fig.~\ref{fig_PD_-1.33}(a). We observe that while the VP phase is stabilized at small twist angles, the IVC phase is preferred at large twist angles. The transition from VP to IVC phase occurs around $\theta\simeq 2.9^\circ$.

First, we focus on the range of lower angles and analyze the valley-polarized states which appear in two slightly different forms labeled VP$_1$ and VP$_2$. The former corresponds to a layer-hybridized situation, while in the latter a non-zero layer-polarization emerges. Figs.~\ref{fig_PD_-1.33}(b) and (c) show the electron density and spin texture in the VP$_1$ and VP$_2$ phases, respectively. The color bar denotes the total charge density and the arrows show the spin direction. Note that the plotted spin component is $ \langle \hat{S}^i_{z} \rangle$, as in the VP phase the in-plane magnetization is homogeneously zero. We find that in both VP$_1$ and VP$_2$ states each site is occupied by one spin-$\uparrow$ electron while the remaining spin-$\downarrow$ electrons are distributed throughout the system. This leads to the maximal possible spin polarization, with the spin-$\uparrow$ sector forming an inert background due to strong ferromagnetic exchange in the system [Eq.~\eqref{eq_Hint}]. In the layer-hybridized VP$_1$ phase, these spin-$\downarrow$ electrons are equally distributed between $A$ and $B$ sublattices, leading to an equal number of doubly occupied sites on both layers [Fig.~\ref{fig_PD_-1.33}(b)]. However, with increasing displacement field due to the appearance of staggered onsite potential [Eq. (\ref{eq_HD})] a charge transfer from one sublattice to the other is generated and the layer polarized state (VP$_2$) becomes stable above some critical value of $\Delta$. As a result in the layer-polarized VP$_2$ phase, all the spin-$\downarrow$ electrons occupy either the $A (\Delta > 0)$ or $B (\Delta < 0)$ sublattice depending on the displacement-field direction. Consequently, in VP$_2$ phase, all the doubly occupied sites reside in one layer [Fig.~\ref{fig_PD_-1.33}(c)] only. 

In both VP$_1$ and VP$_2$ states, a $\sqrt{3}\times\sqrt{3}$ reconstruction is realized, as shown by the unit cells depicted by the dashed line in Figs.~\ref{fig_PD_-1.33} (b) and (c). For the sake of completeness, the insets to the figures show amplitudes of the Fourier transforms (FT) of $(\langle \hat{n}_i \rangle-\bar{n})$ (left inset) and $(\langle \hat{S}^i_z \rangle-\bar{S_z})$ (right inset) inside the mBZ. The mean values are subtracted to remove the large peak at $\textbf{k}=0$, which does not influence other peaks. The occurrences of the peaks at the mBZ corners ($\kappa, \kappa'$) confirm the $\sqrt{3}\times\sqrt{3}$ reconstruction. It should be noted that the maximal spin polarization of the VP phases leads to $\hat{S}^i_z\approx\frac{1}{2}(1-\hat{n}_{i\downarrow})$ at every lattice site, hence, the spin/charge modulation is determined by the active spin-$\downarrow$ sector. As a result, the spin and the charge degrees of freedom are tied to each other and the subtle interplay between the ferromagnetic term, the intersite Coulomb repulsion term, and the staggered potential tuned by the displacement field determine the resulting charge and spin modulation. While the ferromagnetic term maximizes the spin/valley polarization, the Coulomb repulsion terms have a significant influence on the resultant charge ordering while the displacement field additionally initiates polarization between the two layers.

\begin{figure}[t]
    \vspace{6pt}
    \centering
    \includegraphics[width=0.48\textwidth]{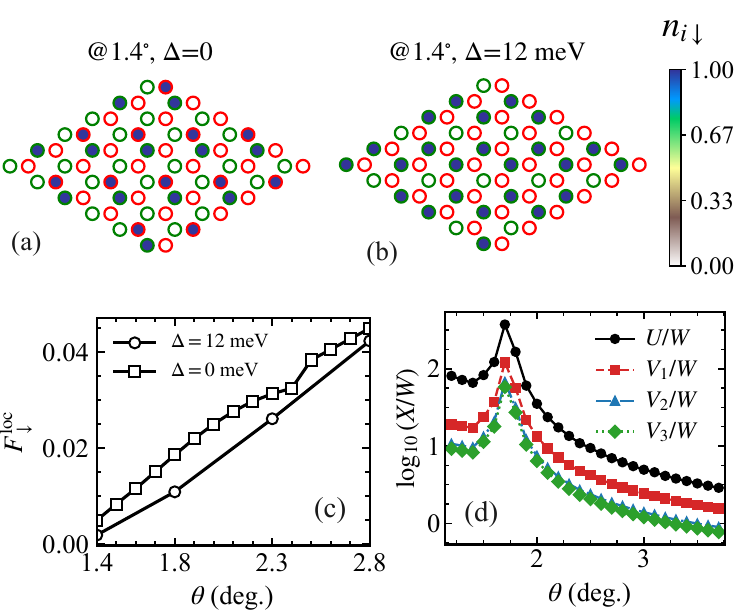}
    \caption{
    Classical nature of VP$_1$ and VP$_2$ phase. (a) Distribution of spin-$\downarrow$ electrons at $\theta=1.4^\circ$ and $\Delta=0$ (VP$_1$ phase). (b) Distribution of spin-$\downarrow$ electrons at $\theta=1.4^\circ$ and $\Delta=12$ meV (VP$_2$ phase). (c) Average local variance of spin-$\downarrow$ electrons as function of $\theta$ at $\Delta=0$ (squares) and $\Delta=12$ meV (circles), i.e. in VP$_1$ and VP$_2$ phase, respectively. (d) Plots the log of the ratios of the Coulomb interaction potentials and the bandwidth (W) of the top \moire valence band as a function of $\theta$. U is the onsite interaction. V$_1$, V$_2$, and V$_3$ denote the interaction between $1^{st}$, $2^{nd}$, and $3^{rd}$ nearest neighbours. $X$ is used as a common notation for onsite and intersite Coulomb potentials.
    }
    \label{fig_VP_classical}
\end{figure}

To further characterize the VP phases, we focus on the spatial distribution of the spin-$\downarrow$ electrons only, as the spin-$\uparrow$ electrons form an essentially inert background. Figs.~\ref{fig_VP_classical}(a) and (b) show the spin-$\downarrow$ electron distributions in VP$_1$ and VP$_2$ phases, respectively. Clearly, the occupation of the sites is near 0 or 1 in both phases. Hence the local variance of spin-$\downarrow$ electrons $\langle \hat{n}_{i\downarrow}\rangle(1-\langle \hat{n}_{i\downarrow}\rangle)$ is strongly suppressed. In the following we examine the average local variance of the spin-$\downarrow$ electrons

\begin{equation}
    F^{loc}_\downarrow=\frac{1}{N_{site}}\sum_i\left(\langle \hat{n}_{i\downarrow}^2\rangle - \langle \hat{n}_{i\downarrow}\rangle^2\right).
\end{equation}
Note that sites with occupation 0 or 1 does not contribute to the variance, whereas fractionally occupied sites produce a finite contribution. For the considered $6\times6$ system, we have 24 spin-$\downarrow$ electrons and the maximum value of $F^{loc}_{\downarrow}$ is $\simeq 0.22$, which is obtained when all these electrons are equally distributed over all the 72 sites. Fig.~\ref{fig_VP_classical}(c) plots $F^{loc}_\downarrow$ as a function of $\theta$ in the VP$_1$ phase ($\Delta=0$) denoted by black squares. We note that $F^{loc}_\downarrow$ remains small in magnitude over the entire twist range. This demonstrates weak delocalization induced by significant electron-electron interactions. Fig.~\ref{fig_VP_classical}(d) plots the log of the interaction-to-bandwidth ratios as a function of twist. Here, the bandwidth ($W$) of the topmost noninteracting \moire valence band is considered, within which the Fermi level of the noninteracting system lies for the hole fillings considered in the study. The large values of the ratios at low twist regime support the dominance of the interaction over the kinetic energy. We further observe that the variance is more suppressed in the VP$_2$ phase as shown by the black circles in Fig.~\ref{fig_VP_classical}(c). This occurs because, due to the finite layer polarization in VP$_2$ phase, the spin-$\downarrow$ electrons are largely restricted to the $A$-sublattice (for $\Delta > 0$) leading to the suppression of the relevant kinetic-energy scale. This consequently enhances the effective interaction-to-bandwidth ratios, resulting in even smaller local charge fluctuations. With an increase of $\theta$, the variance increases, signaling a gradual departure from the classical limit. This behavior is consistent with the decreasing relative importance of interactions at larger twist angles [Fig.~\ref{fig_VP_classical}(d)], which allows stronger charge delocalization and hence larger fluctuations in the local occupations. We therefore conclude that the electrons in the VP phase exhibit weak charge delocalization and the corresponding density distribution resembles classical real-space configurations at zero temperature, where the site occupation is restricted to 0 or 1. However, with increasing twist, the system progressively becomes more quantum in nature.

Next, we move on to the discussion of the IVC phases found at larger $\theta$ in the phase diagram of Fig.~\ref{fig_PD_-1.33}(a). Figs.~\ref{fig_PD_-1.33}(d) and (e) show the electron density and spin texture of the layer-hybridized IVC$_1$ and layer-polarized IVC$_2$, respectively. Since in both IVC states the charge is equally distributed between the two valleys ($\langle \hat{n}_{i\uparrow}\rangle=\langle \hat{n}_{i\downarrow}\rangle$) we have $\langle\hat{S}^z_i\rangle=0$. Therefore, in Figs.~\ref{fig_PD_-1.33}(d) and (e) we plot the in-plane spin component $\langle \hat{\textbf{S}}^i_{\parallel} \rangle=(\langle \hat{S}^i_{x} \rangle, \langle \hat{S}^i_{y} \rangle)$ depicted by the black arrows. In IVC$_1$, the in-plane spins on $A$ and $B$ sublattices are oppositely oriented, thus cancelling each other and giving rise to a sublattice-compensated IVC order, resulting in zero in-plane magnetization. Because of the layer-hybridized nature, in IVC$_1$, the doubly occupied sites are equally distributed on both $A$ and $B$ sublattices. In contrast, in the IVC$_2$ phase, all the doubly occupied sites move to a single sublattice ($A$ for $\Delta > 0$) which is due to the staggered onsite potential induced by the displacement field. As a result, all sites belonging to the sublattice $B$ have a single electron, resulting in a uniform spin in the $B-$sublattice network, while the sublattice $A$ features fewer sites with nonzero spin moment. Therefore, cancellation of in-plane spins between $A$ and $B$ sublattices is incomplete, and a net in-plane spin moment develops. We call this IVC order as ferrimagnetic-IVC. Similar to the VP phases, the IVC phases also show $\sqrt{3}\times\sqrt{3}$ reconstruction in both the charge and spin order. In Figs.~\ref{fig_PD_-1.33}(d) and (e), the reconstructed unit cell is shown by the dashed lines. The left-inset and the right-inset show the amplitudes of the mean-subtracted-FTs of charge and in-plane spin densities, respectively. For in-plane spin at site $i$ we use $S^i_+=S^i_x+iS^i_y$. The peaks in the FT amplitudes occur at the mBZ corners($\kappa, \kappa'$), confirming the  $\sqrt{3}\times\sqrt{3}$ order. 

\begin{figure}[t]
    \centering
    \includegraphics[width=0.49\textwidth]{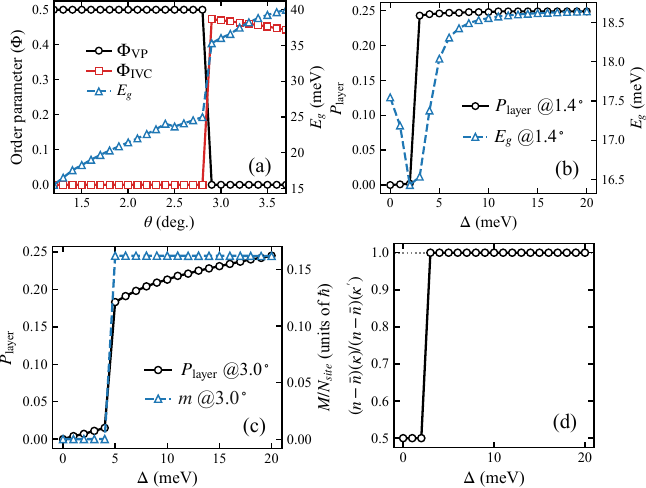} 
    \caption{
    Phase transitions in the 4/3 hole-filling case. (a) VP to IVC phase transition as a function of $\theta$ at $\Delta=0$. The left axis shows the VP (black circles) and the IVC (red squares) order parameters (see text). The right axis shows $E_g$ (blue triangles). (b) Shows $P_{layer}$ (black circles) on the left axis and the $E_g$ (blue triangles) on the right axis for $\theta=1.4^\circ$. (c) Shows $P_{layer}$ (black circles) on the left axis and the magnetization per site (blue triangles) on the right axis for $\theta=3.0^\circ$. The transitions are VP$_1$-VP$_2$ and IVC$_1$-IVC$_2$ for $\theta=1.4^\circ$ and $3.0^\circ$, respectively. For the phases refer to fig.~\ref{fig_PD_-1.33}. (d) The ratio of the amplitudes of the FT peaks of the mean-subtracted charge density at $\kappa$ and $\kappa'$ in the mBZ. This plot holds for any $\theta$; only the location of the jump follows the transition in $P_{layer}$. The amplitudes of the FTs are plotted in the insets of Fig.~\ref{fig_PD_-1.33}.
    }
    \label{fig_PT_-1.33}
\end{figure}

We now focus on the phase transitions among various VP and IVC phases presented in the phase diagram of Fig.~\ref{fig_PD_-1.33}(a). Fig.~\ref{fig_PT_-1.33}(a) characterizes the VP-to-IVC phase transition as a function of $\theta$ at $\Delta=0$. In the following, we define the VP and the IVC order parameters. 
\begin{equation}
    \begin{aligned}
        \Phi_{VP}=\frac{n_\uparrow-n_\downarrow}{n_\uparrow+n_\downarrow},\\
        \Phi_{IVC}=\frac{2}{N_e}\sum_i |S^i_+|.
    \end{aligned}
\end{equation}
Here, $n_\uparrow$, $n_\downarrow$, and $N_e (=n_\uparrow+n_\downarrow)$ are the numbers of spin-$\uparrow$, spin-$\downarrow$, and total electrons, respectively. Across the transition point at $\theta \simeq 2.9^\circ$, the VP order parameter (black circles) shows sudden collapse with concomitant emergence of the IVC order parameter (red squares). The discontinuous jumps in the order parameters reveals the first-order nature of the transition. The single-particle bandgap ($E_g$) at the Fermi level (blue triangles) also shows a sudden jump across the transition. Moving from VP to IVC, the HF Hamiltonian gains valley-off-diagonal terms. These can open additional hybridization gaps resulting in abrupt increase of the bandgap at the transition point. It should be noted that in the considered situation of fractional band filling, the single particle Fermi surfaces do not correspond to nesting vectors which would lead to the observed $\sqrt{3}\times\sqrt{3}$ reconstruction. Also, there is no significant change of the single particle Fermi surface across the transition between the VP and IVC states. Hence, the obtained ordering should not be considered as resulting from the momentum-space instability but rather a real-space phenomenon reflecting the competition between different types of interaction terms and the kinetic energy. In the VP state which is close to the classical limit the interaction terms dominate. However, with increasing twist angle the kinetic energy becomes gradually more important [Fig. \ref{fig_VP_classical} (d)] leading to increased role of fluctuations. At some critical value of $\theta$ a phase transition takes place to the IVC state which is more quantum in nature.

Next, we discuss the influence of the displacement field, which introduces the staggered onsite potential $\Delta$ and leads to layer polarization defined in the following manner
\begin{equation}
    P_{layer}=\frac{|n_A-n_B|}{(n_A+n_B)},
\end{equation}
where $n_A(n_B)$ is the total number of electrons on the $A(B)$ sublattice. Fig.~\ref{fig_PT_-1.33}(b) characterizes the transition from VP$_1$ to VP$_2$ at $\theta=1.4^\circ$. Across the transition at $\Delta \simeq$ 3 meV, $P_{layer}$ (left axis) discontinuously jumps to a nonzero value, indicating the first-order nature of the transition. The $E_g$ (right axis) also shows a rapid change near the transition. Additionally, Fig.~\ref{fig_PT_-1.33}(c) presents the transition from IVC$_1$ to IVC$_2$ state at $\theta=3.0^\circ$. The first-order nature of the transition is evident from the abrupt changes in $P_{layer}$ (left axis). On the right axis we show magnetization per site denoted by $M/N_{site}(=m)$. The jump from zero to finite value of $m$, confirms the emergence of net in-plane magnetization in the IVC$_2$ phase. All the results taken together, we conclude that both VP$_1$-VP$_2$ and IVC$_1$-IVC$_2$ transitions fall in the category of first-order transition. 

Although all the obtained phases in the phase diagram of Fig.~\ref{fig_PD_-1.33}(a) show $\sqrt{3}\times\sqrt{3}$ reconstruction, the layer-polarization transition influences the relative amplitudes of the FT peaks at $\kappa$ and $\kappa'$. In the layer-hybridized regime (VP$_1$ and IVC$_1$) both sublattices ($A$ and $B$) individually contain the $\sqrt{3}\times\sqrt{3}$ order. But in the layer-polarized regime (VP$_2$ and IVC$_2$), one of the sublattices ($B$ for $\Delta > 0$) becomes uniform in terms of the charge and spin densities, and the $\sqrt{3}\times\sqrt{3}$ order is entirely contained in the other layer ($A$ for $\Delta > 0$). As a result, the FTs in layer-hybridized regime show interference between the $A$ and $B$ sublattices, producing unequal amplitudes at $\kappa$ and $\kappa'$, which can be seen in the insets of Figs.~\ref{fig_PD_-1.33}(b) and (d). In contrast, in the layer-polarized regime, the absence of such interference results in equal amplitudes at $\kappa$ and $\kappa'$, as is clear from the insets of Figs.~\ref{fig_PD_-1.33}(c) and (e). Fig.~\ref{fig_PT_-1.33}(d) plots the ratio of the amplitudes for the charge density at $\kappa$ and $\kappa'$ as a function of $\Delta$. The jump in the ratio from 0.5 to 1 coincides with layer-polarization transition. Such a jump is true for any $\theta$; only the location of the jump follows the transition in $P_{layer}$.

\subsection{Hole-filling 3/2: phase diagram, charge, and spin order of the phases}
\begin{figure*}[t]
    \centering
    \includegraphics[width=0.8\textwidth]{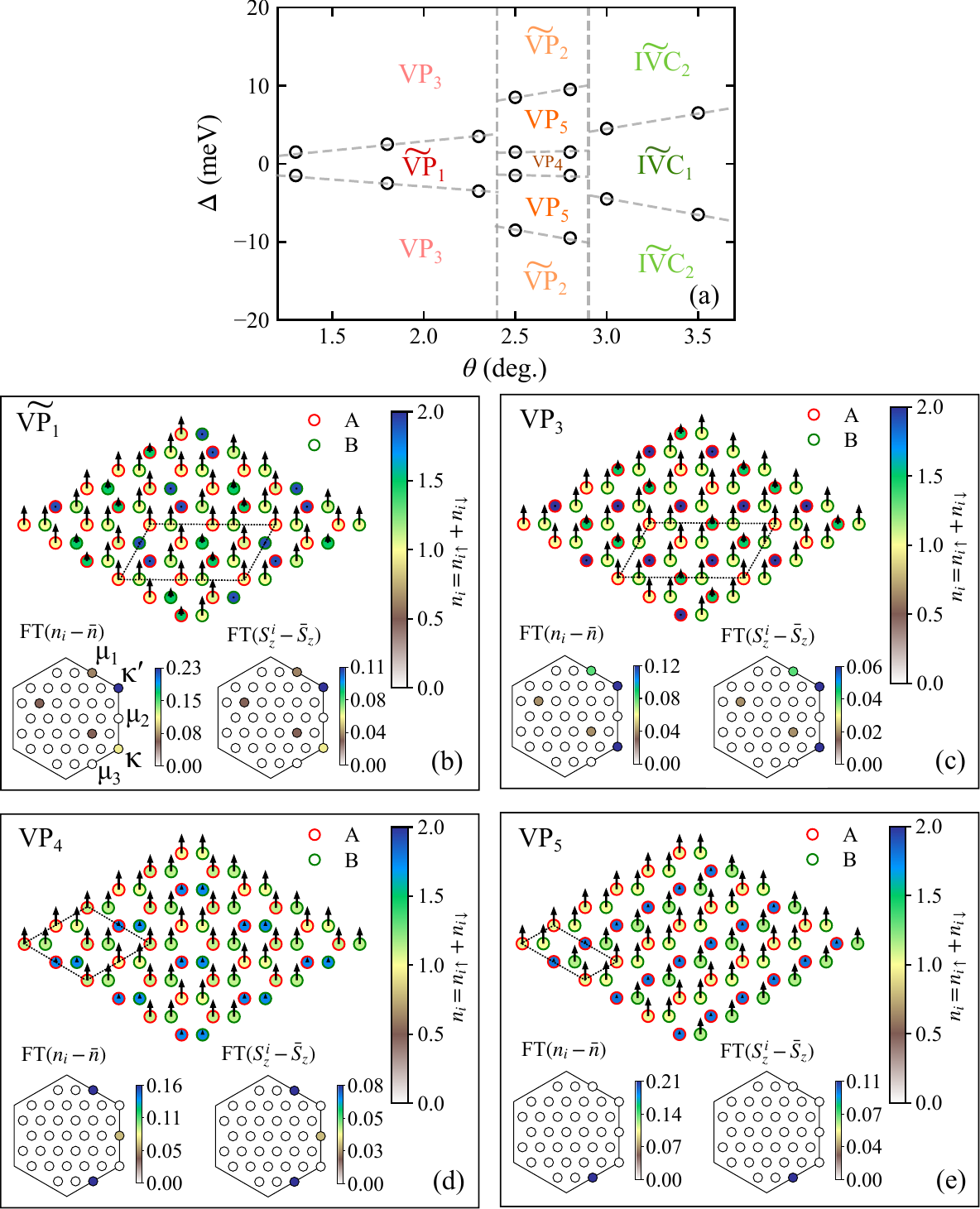}
    \caption{
    Phase diagram of hole-filling 3/2 case. (a) The phase of the ground state as a function of $\theta$ and $\Delta$. The open black circles denote the phase boundary determined from the HF calculations as a function of the displacement field at selected angles. The phases $\widetilde{\mathrm{VP_1}}$, VP$_3$, VP$_4$, and VP$_5$ are characterized in (b), (c), (d), and (e), respectively, for $\Delta > 0$. (b) The 6$\times$6 lattice system is shown by red ($A$ sublattice) and green ($B$ sublattice) circles. The colorbar shows the charge density. The arrows show spin. And the dashed lines depict the reconstructed unitcell. At the bottom, the left inset and the right inset plot the amplitudes of the FT of mean-subtracted charge and spin density, respectively. The details of (c), (d), and (e) are the same as those of (b). Note that the arrows represent the $z$-component of spin.
    }
    \label{fig_PD_-1.50}
\end{figure*}

We now discuss the 3/2 hole-filling case. Fig.~\ref{fig_PD_-1.50}(a) presents the ground state phase diagram as a function of $\theta$ and $\Delta$. This phase diagram shares some similarities with the previous one corresponding to 
filling 4/3 in Fig.~\ref{fig_PD_-1.33}(a). Like the 4/3 filling case, the VP phases possess maximal valley polarization and the VP-IVC transition occurs at $\theta\simeq 2.9^\circ$.  Despite these similarities, the phase diagram of 3/2 case has some new features. A few new orders, $2\times 1$  and $2\times 2$, appear alongside the ubiquitous $\sqrt{3}\times\sqrt{3}$ order as we discuss further on. 

 Compared to the 4/3 hole-filling case, the present case has 1/6 more holes (or equivalently 1/6 fewer electrons) per unit cell. Therefore, the layer-hybridized $\widetilde{\mathrm{VP_1}}$ phase can be understood as a hole-doped version of the VP$_1$ phase of Fig.~\ref{fig_PD_-1.33}(a). In the VP$_1$ phase, the spin-$\downarrow$ electrons are nearly classically localized, with site occupations close to either 0 or 1. Upon increasing the hole filling to 3/2, some of the sites that were nearly fully occupied by spin-$\downarrow$ electrons become approximately half occupied. The additional holes are therefore distributed over twice as many sites, and these half-occupied sites develop a $2\times1$ reconstruction superimposed on the underlying $\sqrt{3}\times\sqrt{3}$ charge order inherited from the VP$_1$ phase, as can be seen in Fig.~\ref{fig_PD_-1.50}(b). The dashed lines show the $2\times1$ reconstruction of the underlying $\sqrt{3}\times\sqrt{3}$ order. Note that the added holes are distributed equally over the $A$ and $B$ sublattices because of the zero layer polarization. The amplitudes of mean-subtracted charge (left inset) and spin density (right inset) FTs show peaks at $\kappa, \kappa'$, and $\mu$ points. The peaks at $\kappa$ and $\kappa'$ originate from $\sqrt{3}\times\sqrt{3}$ ordering and the peak at $\mu$ confirm the further 2x1 reconstruction. 
 Similarly as for the case of 4/3 hole-filling, with an increase in $\Delta$,  the layer-hybridized $\widetilde{\mathrm{VP_1}}$ phase reorganizes into layer-polarized VP$_3$ phase [Fig.~\ref{fig_PD_-1.50}(a)], which is visualized in Fig.~\ref{fig_PD_-1.50}(c). As can see in this phase, all the minority spin-$\downarrow$ electrons are transferred to the $A$ sublattice (for $\Delta > 0$). Therefore, while the $B$ sublattice becomes uniform in terms of the charge and spin density, the actual order is contained in the $A$ sublattice network. In the VP$_3$ phase, 2/3 of the minority spins form a $\sqrt{3}\times\sqrt{3}$ pattern, and the remaining 1/3 form a $2\times1$ reconstruction on top of it. The peak positions of the amplitudes of the mean-subtracted FTs shown in the insets again confirm the mentioned real-space order.

 \begin{figure}[t]
    \vspace{6pt}
    \centering
    \includegraphics[width=0.48\textwidth]{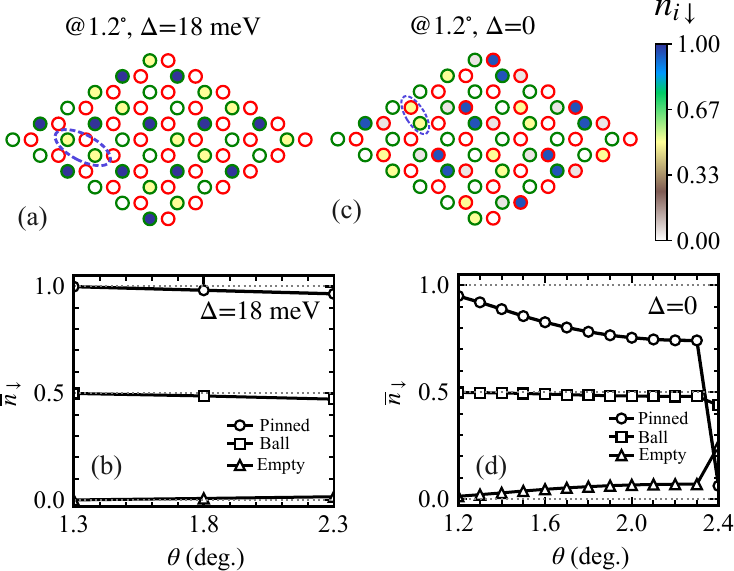}
    \caption{
    Signature of the two-component GWC in $\widetilde{\mathrm{VP_1}}$ and VP$_3$ phase. (a) Distribution of spin-$\downarrow$ electrons at $\theta=1.2^\circ$ and $\Delta=18$ meV (VP$_3$ phase). (b) Average occupation of pinned, ball, and empty sites as a function of $\theta$ at $\Delta=18$ meV (VP$_3$ phase). (c) Distribution of spin-$\downarrow$ electrons at $\theta=1.2^\circ$ and $\Delta=0$ meV ($\widetilde{\mathrm{VP_1}}$ phase). (d) Average occupation of pinned, ball, and empty sites as a function of $\theta$ at $\Delta=0$ ($\widetilde{\mathrm{VP_1}}$ phase). The dashed ovals in (a) and (c) represent the two-site coherent dimers. 
    }
    \label{fig_pinball}
\end{figure}

In the intermediate twist regime, the VP phase reorganizes itself into a genuinely new order rather than a hole-doped version of 4/3 filling. In the layer-hybridized regime a $2\times2$ ordered phase termed as VP$_4$ appears as shown in Fig.~\ref{fig_PD_-1.50}(d). The quadrupling of the original unit cell is indicated by the dashed lines. Similarly as before, the amplitudes of FTs of the mean-subtracted charge (left inset) and spin (right inset) density show peaks at all three $\mu$ points which is consistent with the $2\times2$ real space order. With the increase of $\Delta$, all the spin-$\downarrow$ electrons of VP$_4$ moves to $A$ sublattice and gives rise to $2\times1$ ordered VP$_5$ phase, Fig.~\ref{fig_PD_-1.50}(e). The doubling of the unit cell along one direction gives rise to the peak at a single $\mu$ point in the FT amplitudes in the insets. Further increase of $\Delta$ induces more layer-polarization, and the system goes into $\widetilde{\mathrm{VP_2}}$, Fig.~\ref{fig_PD_-1.50}(a),  which is the hole-doped version of VP$_2$ discussed in connection with Fig.~\ref{fig_PD_-1.33}. This phase shows reconstruction of the underlying $\sqrt3\times\sqrt3$ order (see supplementary material \cite{supple}).

Similar to the 4/3 hole-filling case [Fig.~\ref{fig_PD_-1.33}(a)], in the present case, the system leaves the VP phase and enters the IVC state at $\theta \simeq 2.9^\circ$ [Fig.~\ref{fig_PD_-1.50}(a)]. Both the layer-hybridized $\widetilde{\mathrm{IVC_1}}$ and layer-polarized $\widetilde{\mathrm{IVC_2}}$ show reconstruction of the underlying $\sqrt{3}\times\sqrt{3}$ order. Further discussions of  $\widetilde{\mathrm{IVC_1}}$ and $\widetilde{\mathrm{IVC_2}}$ are relegated to the supplementary material \cite{supple}.

Now we further elaborate on the nature of the $\widetilde{\mathrm{VP}}_1$ and $\mathrm{VP}_3$ phases. These phases arise upon doping holes to $\mathrm{VP}_1$ and $\mathrm{VP}_2$ phases shown in Fig.~\ref{fig_PD_-1.33}(a), in the corresponding ranges of $\theta$ and $\Delta$. As discussed in Fig.~\ref{fig_VP_classical}, the $\mathrm{VP}_1$ and $\mathrm{VP}_2$ phases exhibit a pronounced classical character, with the local electron occupations close to the integer values. However, with additional hole doping, the system departs from this nearly classical limit. Doping holes only reduces the number of spin-$\downarrow$ electrons, as the spin-$\uparrow$ sector remains inert due to maximal spin polarization. Fig.~\ref{fig_pinball}(a) shows the spin-$\downarrow$ occupations in the $\mathrm{VP}_3$ phase at $\theta=1.2^\circ$ and $\Delta=18$ meV. Three distinct groups of sites can be identified, with occupations close to 1, 0.5, and 0. The sites with occupation 1 remain strongly localized with negligible coherence with the nearby sites. Whereas the sites with occupation 0.5, forms dimer like entities. A given half-occupied site shows coherence only with the nearest half-occupied site, ie the electron is confined within these two sites. The dashed oval in Fig.~\ref{fig_pinball}(a) represents such a dimer. The state therefore possesses two qualitatively distinct carrier components: a set of nearly pinned carriers and a set of two-site coherent dimer like charge objects. This is reminiscent of a pinball liquid \cite{Hotta_PRB_2006, Merino_PRL_2013}, in which a subset of carriers remain pinned to sites, while the rest of the carriers with higher mobility form the ball component. However, the $\mathrm{VP}_3$ phase remains gapped as found in HF calculations and is therefore insulating rather than liquid-like. Hence, we refer to this state as a two-component  GWC \cite{Kumar_npj_2025}. Fig.~\ref{fig_pinball}(b) plots the average spin-↓ occupations of the pinned, ball-like, and nearly empty sites as a function of $\theta$. Their occupations vary only weakly with twist angle, indicating that the two-component character remains robust throughout the range in which the $\mathrm{VP}_3$ phase is stable.

Fig.~\ref{fig_pinball}(c) shows the spin-$\downarrow$ occupation pattern in the $\widetilde{\mathrm{VP}}_1$ phase at $\theta=1.2^\circ$ and $\Delta=0$. In contrast to the $\mathrm{VP}_3$ phase, the occupation of the most strongly occupied sites is already reduced from unity, while the nearly empty sites acquire a small but finite occupation. Fig.~\ref{fig_pinball}(d) shows the average occupations of the pinned, ball-like, and nearly empty sites as a function of $\theta$. While the occupation of the ball-like sites remains close to 0.5, we observe a gradual charge transfer from the most highly occupied sites to the least occupied sites as $\theta$ increases. This behavior indicates that the two-component character of the $\widetilde{\mathrm{VP}}_1$ phase weakens rapidly with increasing twist angle. The observed trend further suggests that a more pronounced two-component  GWC may emerge for twist angles below $\theta=1.2^\circ$.

We therefore conclude that, while the parent phases $\mathrm{VP}_1$ and $\mathrm{VP}_2$ are close to classical real-space occupation patterns, upon hole doping the obtained phases $\widetilde{\mathrm{VP}}_1$ and $\mathrm{VP}_3$ display substantially stronger charge delocalization and exhibit signatures of a two-component  GWC.

\begin{figure}[t]
    \centering
    \includegraphics[width=0.49\textwidth]{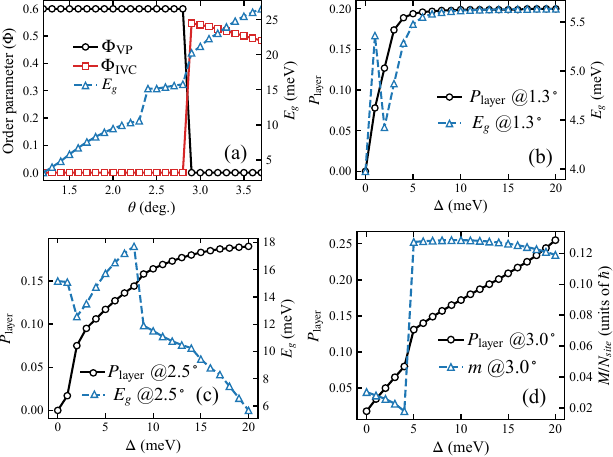}
    \caption{
    Phase transitions in the 3/2 hole-filling case. (a) VP to IVC phase transition as a function of $\theta$ at $\Delta=0$. The left axis shows the VP (black circles) and the IVC (red squares) order parameters. The right axis shows $E_g$ (blue triangles). (b) shows $P_{layer}$ (black circles) on the left axis and the $E_g$ (blue triangles) on the right axis for $\theta=1.3^\circ$. (c) same as (b) for $2.5^\circ$. (d) shows $P_{layer}$ (black circles) on the left axis and magnetization per site (blue triangles) on the right axis for $\theta=3.0^\circ$. For the phases refer to Fig.~\ref{fig_PD_-1.50}.
    }
    \label{fig_PT_-1.50}
\end{figure}

Next, we move on to the characterization of the transitions between the phases shown in Fig.~\ref{fig_PD_-1.50}(a). Namely, Fig.~\ref{fig_PT_-1.50}(a) presents the evolution of the order parameters across the VP-IVC transition induced by tuning $\theta$ at $\Delta=0$. The abrupt changes of the VP (black circles) and IVC (red squares) order parameters which appear at the critical value indicate first order phase transition with increasing angle. In the figure, we also provide the single- particle bandgap at the Fermi level, $E_g$ (blue triangles), which shows two jumps as a function of $\theta$. The first one corresponds to the $\widetilde{\mathrm{VP}}_1$-VP$_4$ transition while the second marks the previously discussed VP-IVC transition. Furthermore, as can be seen from Figs. ~\ref{fig_PT_-1.50}(b) and (c) by increasing $\Delta$ one enhances the layer polarization (black circles) in a straightforward manner, which again leads to abrupt changes in $E_g$ (blue triangles) at the critical values corresponding to the first order phase transitions. It should be noted that the two jumps in $E_g$, which appear for $\theta=2.5^{\circ}$ [Fig.~\ref{fig_PT_-1.50}(c)] correspond to VP$_4$-VP$_5$ and VP$_5$-$\widetilde{\mathrm{VP}}_2$ transitions. For the case of $3^{\circ}$ twist angle, a rapid increase of total in-plane magnetization appears at the transition point, as we show in Fig.~\ref{fig_PT_-1.50}(d).


\subsection{Universal features of the HF phase diagrams}
Here we comment on some general features of the phase diagrams of the 4/3 and 3/2 hole-filling cases. The underlying principles that determine the phases and govern the phase transitions are discussed as well.

1) In both 4/3 and 3/2 hole-filling, while the VP phase stabilizes at low twist angles, the IVC phase is favored at high twist. The transition from VP to IVC occurs at $\theta\simeq2.9^\circ$ for both fillings. The VP phase is exchange-driven, originating from the term $-2J_{ij}\hat{S}^z_i\hat{S}^z_j$ in Eq.~\eqref{eq_Hint}. This transition  reflects a reorganization of the HF state in which the energy gain from intervalley coherence becomes larger than the easy-axis exchange energy of the VP state. Once the IVC phase appears, the HF Hamiltonian gains valley-off-diagonal terms. These can lead to an additional hybridization gap leading to a decrease in the total energy of the system. This explains the jumps in $E_g$ across the VP to IVC transitions in Fig.~\ref{fig_PT_-1.33}(a) and Fig.~\ref{fig_PT_-1.50}(a).

2) In the VP phase, the spin polarization is completely saturated. The majority spin occupying each site acts as an inert background. The minority spin electrons are the actual degrees of freedom, whose reorganization determines the charge order of the phase. This also explains why the spin order is tied to the charge order in the VP phase.

The phase diagrams reveal the ubiquity of the $\sqrt{3}\times\sqrt{3}$ order. For hole-filling 4/3, the entire phase diagram is occupied by $\sqrt{3}\times\sqrt{3}$ reconstructions irrespective of the VP or IVC phase. Similarly, for hole-filling 3/2, the same order appears everywhere except in the intermediate range of $\theta$-$\Delta$. The reason behind this universality is twofold: commensurability and triangular lattice formation. It is well known that through triangular lattice formation, which leades to $\sqrt{3}\times\sqrt{3}$ reconstruction, interacting particles maintain equal and maximum distances from each other, resulting in a minimum interaction energy. For hole-filling 4/3, commensurability further reinforces the order. In this case, the average number of electrons per unit cell is 8/3, which is commensurable with the $\sqrt{3}\times\sqrt{3}$ order. So both triangular lattice formation and commensurability work in favor of $\sqrt{3}\times\sqrt{3}$ reconstruction. Unlike hole-filling 4/3, the extra dimension of commensurability is lost to some extent for the hole-filling 3/2 case, as the average number of electrons per unit cell is 5/2. Equivalently, each unit cell has 1/6 extra holes compared to the 4/3 hole-filling case. These extra holes are accommodated appropriately through a $2\times1$ or $2\times2$ reconstruction of already existing $\sqrt{3}\times\sqrt{3}$ order. 

So we see that $\sqrt{3}\times\sqrt{3}$ is favored by the energy minimization principle and is further reinforced by the commensurability condition. These underlying principles that govern the order of the phases are independent of the size of the system. This tells us that the phases and phase diagrams obtained in our work for the finite $6\times6$ system will also hold for larger system sizes.

\section{Variational Monte Carlo Results}
We have further employed the VMC technique to study the effects of the interactions beyond the mean-field considerations. We performed VMC simulations using both random initialization of variational parameters and initialization with the results of the converged solutions of the unrestricted HF calculations. In general, the lowest optimized energy is obtained when the converged HF results are used for initialization, which we report here. The simulations confirm the phases obtained with the HF method for both 4/3 and 3/2 hole-filling cases. In the paper, we primarily discuss the results of 4/3 hole-filling case.

\begin{figure}[t]
    \vspace{6pt}
    \centering
    \includegraphics[width=0.48\textwidth]{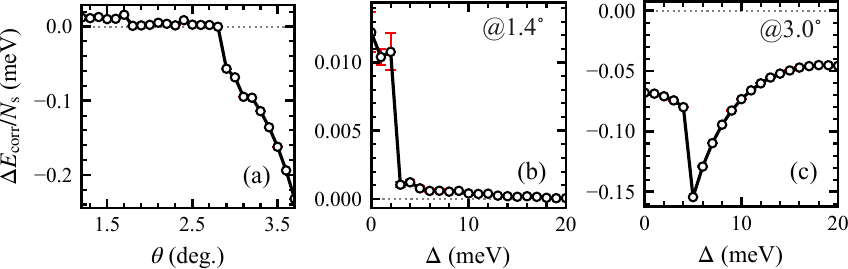}
    \caption{ Correlation energy for 4/3 hole-filling case.
    (a) Correlation energy per site as function of $\theta$ at $\Delta=0$. (b) and (c) show the correlation energies per site as function of $\Delta$ for $\theta=1.4^\circ$ and $3.0^\circ$, respectively.
    }
    \label{fig_corr_energy}
\end{figure}

Although the VMC simulations produced ordered phases similar to those obtained within the HF calculations, the ground-state energies obtained in the two methods differ in some parameter range.  The importance of correlations beyond the mean-field level can be directly assessed by comparing the ground state energies obtained from the two methods. For this purpose, we study the correlation energy defined in the following manner, 
\begin{equation}
 \Delta E_{\mathrm{corr}}=E_{\mathrm{VMC}}-E_{\mathrm{HF}},
\end{equation}
where $E_{\mathrm{VMC}}$ and $E_{\mathrm{HF}}$ denote the total energies of the system obtained in the VMC and the HF simulations, respectively. A negative value of $\Delta E_{\mathrm{corr}}$ therefore indicates an energy gain induced by correlations beyond mean-field. Fig.~\ref{fig_corr_energy} (a) shows $\Delta E_{\mathrm{corr}}$ per site as a function of $\theta$ at $\Delta=0$. In the range of $1.2^\circ-2.8^\circ$, where the ground state phase is  VP$_1$, $E_{\mathrm{VMC}}$ agrees very well with $E_{\mathrm{HF}}$. The small positive value of $\Delta E_{\mathrm{corr}}$ arises from the statistical uncertainty of the VMC simulations. This result is consistent with the nearly classical character of the VP$_1$ phase arising from high interaction-to-bandwidth ratio. The corresponding localized charge configuration is captured well by the unrestricted HF solution, leaving only a very small room for improvement by the VMC wave function. With increasing $\theta$, the bandwidth increases relative to the interaction scale, enhancing charge delocalization. At $\theta\simeq2.9^\circ$, the system enters the IVC$_1$ phase, characterized by a coherent superposition of the two spin (valley) degrees of freedom. In this regime, $E_{\mathrm{VMC}}$ becomes lower than $E_{\mathrm{HF}}$, indicating a gain in correlation energy as can be seen in Fig.~\ref{fig_corr_energy}(a). We further note that the gain rapidly increases with $\theta$, which implies growing importance of correlations beyond HF in the IVC phase. 

We previously noted that a sufficiently strong displacement field induces finite layer polarization. The increased layer polarization suppresses interlayer hybridization leading to an enhanced charge localization. Consequently, the energetic importance of fluctuations beyond HF decreases in the layer-polarized regime. Fig.~\ref{fig_corr_energy}(b) plots the correlation energy at $\theta=1.4^\circ$ as function of $\Delta$. Since the VP phase is strongly classical in nature, HF and VMC energies agree and the layer-polarization induces no further effect. The small positive value is due to the statistical uncertainty of the VMC procedure. Whereas in the IVC phase at $\theta=3.0^\circ$ in Fig.~\ref{fig_corr_energy}(c), the reduced fluctuation in the IVC$_2$ phase at higher $\Delta$ results in lower $\Delta E_{\mathrm{corr}}$ as a result $E_{\mathrm{VMC}}$ approaches $E_{\mathrm{HF}}$. Therefore, we conclude that the gain in correlation energy is progressively suppressed as the system evolves toward the layer-polarized state.

Next, we study how rigid the obtained ordered phases are within both HF and VMC approaches. We note that, since no translational-symmetry restoration is imposed during the optimization of the variational wavefunction, the final optimized VMC states can retain the broken translational symmetry. Consequently, the static order of the converged phases can be extracted from the single-particle Greens functions like local charge ($n_i$) and spin ($\textbf{S}_i$) densities. All phases obtained for the 4/3 hole-filling case show  $\sqrt{3}\times\sqrt{3}$ order, which leads to dominant peaks at the mBZ corners ($\kappa, \kappa'$) in the FT of the density modulations. Therefore, we define the strength of the static charge and spin orders order as 
\begin{equation}
\begin{aligned}
    m_\rho=|n(\kappa)|^2+|n(\kappa')|^2, \\
    m_{S_z}=|S_z(\kappa)|^2+|S_z(\kappa')|^2, \\
    m_{S_+}=|S_+(\kappa)|^2+|S_+(\kappa')|^2.     
\end{aligned}    
\end{equation}
To quantify the fluctuation around the mean-field order, we consider the connected 2-particle correlation functions $N^{conn}(\textbf{q}), S^{conn}_{zz}(\textbf{q})$, and $S^{conn}_{+-}(\textbf{q})$, which characterize the charge, out-of-plane spin, and in-plane spin fluctuations, respectively. These quantities measure the strength of the fluctuation of the respective order at wavevector $\textbf{q}$ and are defined as follows
\begin{equation}
\begin{aligned}
N^{conn}(\textbf{q}) &= \frac{1}{N_{site}}\sum_{i,j} e^{i\textbf{q}.(\textbf{r}_i-\textbf{r}_j)}\left(\langle\hat{n}_i\hat{n}_j\rangle -\langle \hat{n}_i\rangle \langle \hat{n}_j\rangle\right),\\
S^{conn}_{zz}(\textbf{q}) &= \frac{1}{N_{site}}\sum_{i,j} e^{i\textbf{q}.(\textbf{r}_i-\textbf{r}_j)}\left(\langle\hat{S}^{i}_z\hat{S}^j_z\rangle-\langle\hat{S}^{i}_z\rangle\langle\hat{S}^j_z\rangle \right),\\
S^{conn}_{+-}(\textbf{q}) &= \frac{1}{N_{site}}\sum_{i,j} e^{i\textbf{q}.(\textbf{r}_i-\textbf{r}_j)}\left(\langle\hat{S}^{i}_+\hat{S}^j_-\rangle-\langle\hat{S}^{i}_+\rangle\langle\hat{S}^j_-\rangle \right).
\end{aligned}
\label{eq:xy}
\end{equation}

We define the total fluctuation by summing the  connected 2-particle correlation functions over all wavevector modes, which is essentially the total local variance of the corresponding quantity. Moreover, we define the order-fluctuation ratio (OFR), denoted by $\eta$, as follows
\begin{equation}
    \begin{aligned}
        \eta_\rho=\frac{m_\rho}{\frac{1}{N_{site}}\sum_\textbf{q}N^{conn}(\textbf{q})+m_\rho},\\
        \eta_{S_z}=\frac{m_{S_z}}{\frac{1}{N_{site}}\sum_\textbf{q}S^{conn}_{zz}(\textbf{q})+m_{S_z}},\\
        \eta_{S_+}=\frac{m_{S_+}}{\frac{1}{N_{site}}\sum_\textbf{q}S^{conn}_{+-}(\textbf{q})+m_{S_+}}.
    \end{aligned}
\end{equation}
\begin{figure}[t]
    \vspace{6pt}
    \centering
    \includegraphics[width=0.48\textwidth]{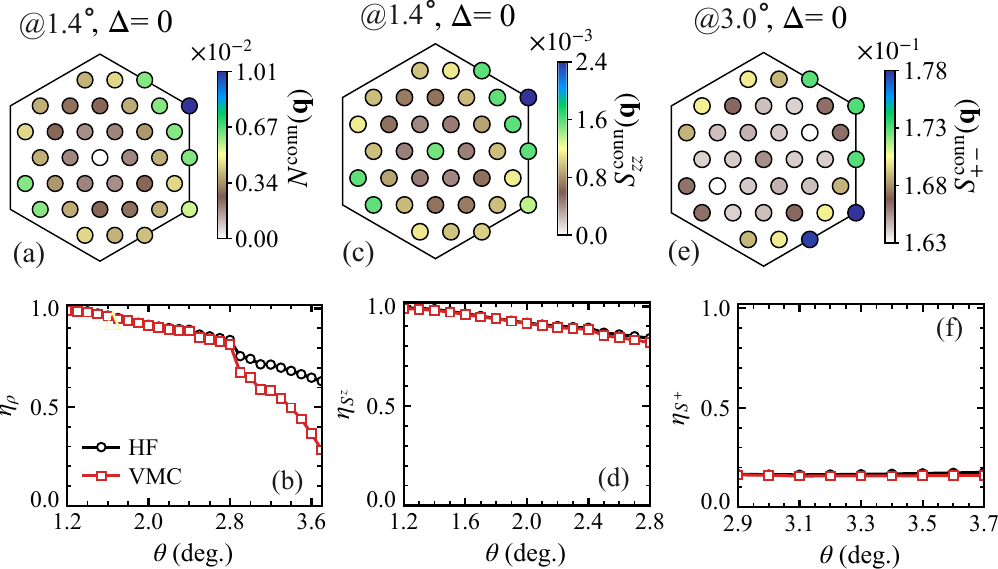}
    \caption{
    OFR of the charge and spin orders for 4/3 hole-filling case as function of twist at $\Delta=0$. (a) $N^{conn}(\textbf{q})$ at $\theta=1.4^\circ$. (b) Charge OFR $\eta_\rho$ as a function of $\theta$. (c) $S^{conn}_{zz}(\textbf{q})$ at $\theta=1.4^\circ$. (d) OFR of the out-of-plane spin order $\eta_{S^z}$ in the VP$_1$ phase. (e) $S^{conn}_{+-}(\textbf{q})$ at $\theta=3.0^\circ$. (f) OFR of the in-plane spin order $\eta_{S^+}$ in the IVC$_1$ phase. In (b), (d), and (f) HF and VMC results are denoted by black circles and red squares, respectively.
    }
    \label{fig_rigidity_Delta=0}
\end{figure} 
As can be seen from the equations above, the OFRs compare the strength of the static order and the fluctuations present in a given phase. The parameter $\eta$ is defined to take values between 0 and 1, where $\eta\approx1$ represents a phase dominated by static-order, while $\eta\approx 0$ implies a phase dominated by fluctuations.

Analysis of the HF and VMC results show that fluctuations of the charge order are generally the strongest at the ordering wavevector, i.e., at $\textbf{q}=\kappa, \kappa'$, throughout the entire twist and displacement field range studied here. Fig.~\ref{fig_rigidity_Delta=0}(a) shows the $N^{conn}(\textbf{q})$ at $\theta=1.4^\circ$ and $\Delta=0$ in the mBZ clearly revealing the pronounced peak at $\kappa'$. The corresponding charge OFR as a function of $\theta$ is shown in Fig.~\ref{fig_rigidity_Delta=0}(b). Within the VP$_1$ phase, which is stable over $1.2^\circ \leq \theta \leq2.8^\circ$, the charge order shows high values of OFR both within the HF and VMC methods, implying a static order dominated crystal. However, with increasing angle, the ordering gradually softens. Upon entering the IVC$_1$ phase, $\theta \geq2.9^\circ$, the two methods lead to significantly different behavior, with the VMC showing a much lower OFR or equivalently softer crystal dominated by fluctuations. This result again shows the highly classical character of the VP$_1$ phase and the increased quantum nature of the IVC phase, where the importance of correlations beyond HF leads to stronger fluctuations.

Similar to charge order, the dominant fluctuation of the out-of-plane spin order also occurs at the ordering wavevectors. Fig.~\ref{fig_rigidity_Delta=0}(c) shows $S^{conn}_{zz}(\textbf{q})$ at $\theta=1.4^\circ$ and $\Delta=0$, where the peak is clearly seen at $\kappa'$.  The $\textbf{q}=0$ component is proportional to the variance of the total out-of-plane spin, $S^{conn}_{zz}(\textbf{q}=0)=\frac{1}{N_{site}}(\langle(\hat{S}^{tot}_z)^2\rangle-\langle\hat{S}^{tot}_z\rangle^2)$, where $\hat{S}^{tot}_z=\sum_i \hat{S}^i_z$. Its small magnitude therefore reflects suppressed fluctuations of $\hat{S}^{tot}_z$, indicating that the ground state is close to an eigenfunction of the $\hat{S}^{tot}_z$. Fig.~\ref{fig_rigidity_Delta=0}(d) plots the OFR of the out-of-plane spin order at $\Delta=0$ over the stability range of the VP$_1$ phase. Like the charge OFR of Fig.~\ref{fig_rigidity_Delta=0}(b), HF (black circles) and VMC (red squares) results are in close agreement. The order is highly rigid at low twist and gradually becomes softer with $\theta$. This effect is induced by the reduced interaction-to-bandwidth ratio, which leads to higher fluctuations around the static order for higher angles.

The character of the fluctuations of the in-plane spin order is qualitatively very different. $S^{conn}_{+-}(\textbf{q})$ at $\theta=3.0^\circ$ and $\Delta=0$ is nearly momentum-independent, as shown in Fig.~\ref{fig_rigidity_Delta=0}(e), indicating that the fluctuations are predominantly short-ranged in real space and lack any preferred ordering wavevector. The weak enhancement at $\kappa$ only reflects the underlying $\sqrt3\times\sqrt3$ modulation. Fig.~\ref{fig_rigidity_Delta=0}(f) plots the corresponding OFR in the twist range where the IVC$_1$ is stable at $\Delta=0$. The small value of OFR implies that in-plane spin order is intrinsically soft, due to the stronger fluctuations present in the IVC phase.

\begin{figure}[t]
    \vspace{6pt}
    \centering
    \includegraphics[width=0.48\textwidth]{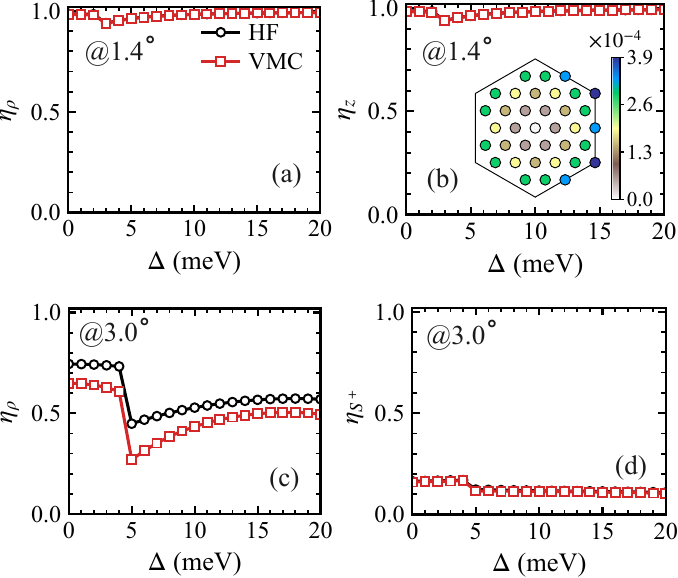}
    \caption{
    OFR of the charge and spin orders for 4/3 hole-filling case at $\Delta\neq0$. (a) Charge OFR at $\theta=1.4^\circ$. (b) Out-of-plane spin OFR at $\theta=1.4^\circ$. The inset shows $S^{conn}_{zz}(\textbf{q})$ at $\Delta=15$ meV (VP$_2$ phase). (c) Charge OFR at $\theta=3.0^\circ$. (d) In-plane spin OFR at $\theta=3.0^\circ$. In (a), (b), (c), and (d) the black circles and the red squares denote the HF and VMC results, respectively.
    }
    \label{fig_rigidity_Delta}
\end{figure} 

For the sake of completeness, we have also analyzed the influence of the layer polarization on OFR. According to our study, with increasing $\Delta$ the system generally maintains high rigidity in the OFR for both charge and out-of-plane spin orders in the VP phase, barring brief dips near $\Delta \simeq3$ meV caused by heightened fluctuations near phase transitions as shown in Fig.~\ref{fig_rigidity_Delta} (a) and (b), respectively. The inset in Fig.~\ref{fig_rigidity_Delta}(b) shows $S^{conn}_{zz}(\textbf{q})$ at $\Delta=15$ meV. A comparison with Fig.~\ref{fig_rigidity_Delta=0}(c), reveals that layer-polarization reduces the fluctuation by almost an order. The vanishingly small value at $\textbf{q}=0$, indicates negligible variance of $\hat{S}^{tot}_z$, essentially making VP$_2$ an eigenstate of $\hat{S}^{tot}_z$. In contrast, the charge OFR in the IVC phase experiences stronger fluctuations and a sharp drop at the layer polarization transition, recovering only partially at higher displacement fields due to suppressed interlayer hybridization as shown in Fig.~\ref{fig_rigidity_Delta} (c). However, the in-plane spin order in the IVC phase remains intrinsically soft, exhibiting a very low OFR across the entire displacement field range as shown in Fig.~\ref{fig_rigidity_Delta}(d).

We finally investigate the origin of the correlation energy gain in the IVC phase discussed in connection with Fig.~\ref{fig_corr_energy}. As mentioned in the Model and Method section, the variational wavefunction, $|\Psi_T\rangle=\hat{\mathcal{P_G}} \hat{\mathcal{P_J}}|\Psi_0\rangle$, used in our VMC study has several variational components and therefore the correlations beyond HF can be incorporated at different levels. First, the pair-product wavefunction, $|\Psi_0\rangle$, itself already spans a more general variational space than the single-determinant wavefunction of HF. Consequently, even in the absence of the Gutzwiller and Jastrow projectors, optimization of pair amplitudes $f_{i\sigma,j\sigma'}$ alone could lead to a solution which is closer to the true ground state of the system. The Gutzwiller and Jastrow projectors can provide additional energy gain by incorporating further many-body correlations into the variational state.

To disentangle the contributions of the different components of the variational wave function, we performed a series of VMC simulations in which different subsets of the variational parameters were optimized. In addition, we performed simulations where the pair amplitudes $f_{i\sigma,j\sigma'}$ are constructed from the converged solutions of the HF simulations and are held fixed, denoted by $f_{HF}$, while the Gutzwiller ($g$) and/or Jastrow ($v$) parameters are optimized. Comparison among these calculations allows us to determine the importance of different variational parameters in the IVC phase of the system.

\begin{figure}[t]
    \vspace{6pt}
    \centering
    \includegraphics[width=0.48\textwidth]{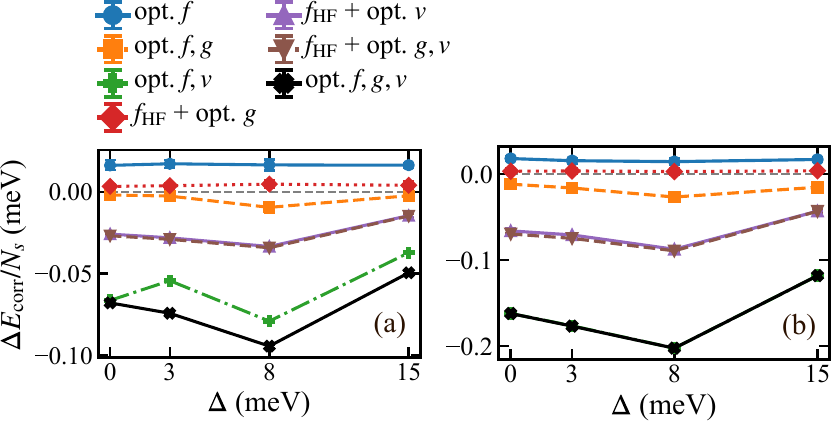}
    \caption{
    Correlation energy per site obtained by optimizing different subsets of the variational parameters at $\Delta=$ 0, 3 meV, 8 meV, and 15 meV for $\theta=3.0^\circ$ (a) and $3.5^\circ$ (b). Only $f_{i\sigma,j\sigma'}$ is optimized: blue circles. Simultaneous optimization of $f_{i\sigma,j\sigma'}$ and $g_i$: orange squares. Simultaneous optimization of $f_{i\sigma,j\sigma'}$ and $v_{ij}$: green pluses. Keeping $f_{i\sigma,j\sigma'}$ fixed to $f_{HF}$ we optimize only $g_i$ (red diamonds), only $v_{ij}$ (purple uparrows), $g_{i}$ and $v_{ij}$ together (brown downarrows). Simultaneous optimization of all parameters $f_{i\sigma,j\sigma'}$, $g_i$, and $v_{ij}$ is denoted by black crosses.
    }
    \label{fig_fgv_opt}
\end{figure}

Figs.~\ref{fig_fgv_opt}(a) and (b) plot the correlation energies obtained by optimizing various subsets of parameters for $\theta=3.0^\circ$ and $3.5^\circ$, respectively, for four different displacement fields. As one can see, the results of the simultaneous optimization of all variational parameters, $f_{i\sigma,j\sigma'}$, $g_i$, and $v_{ij}$, denoted by the black crosses, lead to the smallest energies. As expected, such a procedure yields the largest overall correlation-energy gain. Among all the other scenarios, the one that corresponds to the simultaneous optimization of the pair-product wave function ($f_{i\sigma,j\sigma'}$) and Jastrow parameters ($v_{ij}$), denoted by the green plus signs, is the closest to the fully optimized state. For $\theta=3.0^\circ$ the former solution is slightly above the latter in energy, which demonstrates a small but finite additional contribution from the Gutzwiller sector when all variational degrees of freedom are allowed to readjust simultaneously. For the case of $\theta=3.5^\circ$ the two solutions give almost identical energies.

We conclude from the results presented in Fig.~\ref{fig_fgv_opt} that the gain in the correlation energy in the IVC phase is dominated by the intersite correlations captured by the Jastrow parameters. In contrast, the Gutzwiller projector, which primarily modifies local correlations associated with the double occupancy, plays a minor role in improving the energy gain. This is consistent with the fact that for the particular band filling studied here, the number of electrons exceeds the number of sites, and, as a result, the double occupancy is unavoidable.

\section{Summary}
In this work, we have investigated the charge and spin order of the twisted MoTe$_2$ bilayer system at hole fillings 4/3 and 3/2, which are experimentally found to be insulating. Using an interacting two-band tight-binding model, we construct the phase diagram as a function of the twist angle and an out-of-plane displacement field within the HF approximation. We further examined the effects of correlations beyond the mean-field using the VMC method.

For both fillings, low twist angles stabilize a VP phase with fully saturated out-of-plane ferromagnetism driven by easy-axis ferromagnetic exchange. At $4/3$ filling, the VP phase exhibits a strongly classical character due to a high interaction-to-bandwidth ratio, which is further enhanced by displacement-field-induced layer polarization. In contrast, the VP phase at $3/2$ filling displays a more quantum nature, forming a two-component generalized Wigner crystal that is particularly robust in the layer-polarized regime.

Above $\theta \simeq 2.9^\circ$ a transition from the VP to an IVC state appears for both fillings. In the layer-hybridized regime, opposing in-plane spin moments on the $A$ and $B$ sublattices cancel out, whereas layer polarization induces incomplete cancellation and a net ferrimagnetic in-plane magnetization. We found that, while $4/3$  filling universally favors $\sqrt{3}\times\sqrt{3}$ order due to commensurability and Coulomb repulsion, $3/2$ filling accommodates excess holes through superimposed $2\times 1$ and $2\times 2$ reconstructions.

In the second part of our analysis we have applied the VMC method for the case of 4/3 filling to study the effects of correlations beyond HF. The phases obtained in the VMC simulations remain consistent with those found in the HF simulations. The ground state energies obtained in both methods agree well in the VP phase, which is consistent with its strong classical nature. However, VMC yields noticeably lower energy in the IVC phase. By selectively optimizing different components of the variational wavefunction, we find that this gain in correlation energy is dominated by the nonlocal density correlations, which are captured by the Jastrow projector. The contribution from the local density correlations described by the Gutzwiller factor is rather small.

The rigidity analysis provides a further understanding of the obtained phases. Both charge and spin orders show high rigidity in the low-twist VP phase. In contrast, the high-twist IVC phase is fluctuation-dominated, which leads to substantially lower rigidity. In particular, the in-plane spin order in the IVC phase is very soft. The VMC also predicts a much lower charge rigidity in this regime. The pronounced softening of the charge order at higher twist angle indicates a possible melting of the insulating charge order. Whether the high fluctuations in the IVC states ultimately lead to the metallic state is left for future study. 

Overall, our study reveals a transition from an interaction-dominated classical generalized Wigner crystal at low twist angles to an increasingly fluctuation-dominated quantum state at high twist angles. The results further demonstrate that correlations beyond the mean-field approximation are primarily nonlocal Jastrow-type and are operative only in the quantum phase at high twist angles.

\section{ACKNOWLEDGMENT}
This research was funded by National Science Centre, Poland (NCN) according to Decision No. 2021/42/E/ST3/00128. This work was partly supported by the AGH University of Krakow subvention. We gratefully acknowledge Poland’s high-performance Infrastructure PLGrid ACK Cyfronet AGH for providing computer facilities and support within computational grant no. PLG/2026/019274.


\begin{thebibliography}{99}
%
\bibitem{Wigner_PR_1934}
E. Wigner, Phys. Rev. \textbf{46}, 1002 (1934).
%
\bibitem{Grimes_PRL_1979}
C. C. Grimes and G. Adams, Phys. Rev. Lett. \textbf{42}, 795 (1979).
%
\bibitem{Andrei_PRL_1988}
E. Y. Andrei, G. Deville, D. C. Glattli, F. I. B. Williams, E. Paris, and B. Etienne, Phys. Rev. Lett. \textbf{60}, 2765 (1988).
%
\bibitem{Tsui_Nature_2024}
Y. C. Tsui, M. He, Y. Hu, E. Lake, T. Wang, K. Watanabe, T. Taniguchi, M. P. Zaletel, and A. Yazdani, Nature \textbf{628}, 287 (2024).
%
\bibitem{Smolenski_Nature_2021}
T. Smoleński, P. E. Dolgirev, C. Kuhlenkamp, A. Popert, Y. Shimazaki, P. Back, X. Lu, M. Kroner, K. Watanabe, T. Taniguchi, I. Esterlis, E. Demler, and A. Imamoğlu, Nature \textbf{595}, 53–57 (2021).
%
\bibitem{Zhou_Nature_2021}
Y. Zhou, J. Sung, E. Brutschea, I. Esterlis, Y. Wang, G. Scuri, R. J. Gelly, H. Heo, T. Taniguchi, K. Watanabe, G. Zaránd, M. D. Lukin, P. Kim, E. Demler, and H. Park, Nature \textbf{595}, 48–52 (2021).
%
\bibitem{Xu_Nature_2020}
Y. Xu, S. Liu, D. A. Rhodes, K. Watanabe, T. Taniguchi, J. Hone, V. Elser, K. F. Mak, and J. Shan, Nature \textbf{587}, 214–218 (2020).
%
\bibitem{HLi_Nature_2021}
H. Li, S. Li, E. C. Regan, D. Wang, W. Zhao, S. Kahn, K. Yumigeta, M. Blei, T. Taniguchi, K. Watanabe, S. Tongay, A. Zettl, M. F. Crommie, and F. Wang, Nature \textbf{597}, 650–654 (2021).
%
\bibitem{XHuang_NatPhys_2021}
X. Huang, T. Wang, S. Miao, C. Wang, Z. Li, Z. Lian, T. Taniguchi, K. Watanabe, S. Okamoto, D. Xiao, S. F. Shi, Y. T. Cui, Nat. Phys. \textbf{17}, 715–719 (2021).
%
\bibitem{Li_Science_2024}
H. Li, Z. Xiang, A. P. Reddy, T. Devakul, R. Sailus, R. Banerjee, T. Taniguchi, K. Watanabe, S. Tongay, A. Zettl, L. Fu, M. F. Crommie, and F. Wang, Science \textbf{385}, 86 (2024).
%
\bibitem{Padhi_PRB_2021}
B. Padhi, R. Chitra, and P. W. Phillips, Phys. Rev. B \textbf{103}, 125146 (2021).
%
\bibitem{Pan_PRB_2020}
H. Pan, F. Wu, and S. Das Sarma, Phys. Rev. B \textbf{102}, 201104(R) (2020).
%
\bibitem{Matty_NatCommun_2022}
M. Matty and E.-A. Kim, Nat. Commun. \textbf{13}, 7098 (2022).
%
\bibitem{Duran_PRB_2023}
N. Morales-Durán, P. Potasz, and A. H. MacDonald, Phys. Rev. B \textbf{107}, 235131 (2023).
%
\bibitem{Ung_PRB_2023}
S. F. Ung, J. Lee, and D. R. Reichman, Phys. Rev. B \textbf{108}, 245113 (2023).
%
\bibitem{Tsang_PRR_2023}
Y. Tan, P. K. H. Tsang, V. Dobrosavljević, and L. Rademaker, Phys. Rev. Res. \textbf{5}, 043190 (2023).
%
\bibitem{ZSong_PRB_2023}
Z. Song, U. F. P. Seifert, Z.-X. Luo, and L. Balents, Phys. Rev. B \textbf{108}, 155109 (2023).
%
\bibitem{YZhou_PRL_2024}
Y. Zhou, D. N. Sheng, and E.-A. Kim, Phys. Rev. Lett. \textbf{133}, 156501 (2024).
%
\bibitem{Erkensten_PRB_2024}
D. Erkensten, S. Brem, R. Perea-Causin, and E. Malic, Phys. Rev. B \textbf{110}, 155132 (2024).
%
\bibitem{Kumar_npj_2025}
A. Kumar, C. Lewandowski, and H. J. Changlani, npj Quantum Mater. \textbf{10}, 95 (2025).
%
\bibitem{Biborski_PRB_2025}
A. Biborski and M. Zegrodnik, Phys. Rev. B \textbf{111}, 075116 (2025)
%
\bibitem{Biborski_Polonica_2026}
A. Biborski, Acta. Phys. Pol. B \textbf{57}, 5-A17 (2026).
%
\bibitem{Mishra_arxiv_2026}
S. Mishra, T. M. R. Wolf, and A. H. MacDonald, arXiv:2603.05714 (2026).
%
\bibitem{YZhou_arxiv_2025}
Y. Zhou, I. Esterlis, T. Smoleński, arXiv:2509.21222 (2025).
%
\bibitem{Saha_PRB_2025}
P. Saha, L. Rademaker, and M. Zegrodnik, Phys. Rev. B \textbf{112}, 045147 (2025).
%
\bibitem{Peng_PRB_2026}
L. Peng, C. De Beule, Y. Lai, D. Li, L. Yang, E. J. Mele, S. Adam, Phys. Rev. B \textbf{113}, L161119 (2026).
%
\bibitem{Chen_SciBull_2026}
J. Chen, Q. Li, X. Wang, W. Li, Sci. Bull. \textbf{71}, 1034–1042 (2026).
%
\bibitem{Wang_PRB_2026}
B. Wang, J. Yu, and Cheng-Cheng Liu, Phys. Rev. B \textbf{113}, L201409 (2026).
%
%
\bibitem{Hubbard_PRB_1978}
J. Hubbard, Phys. Rev. B \textbf{17}, 494 (1978).
%
\bibitem{Noda_PRL_2002}
Y. Noda, M. Imada, Phys. Rev. Lett. \textbf{89}, 176803 (2002).
%
\bibitem{Gruner_1994}
G. Grüner (1994). Density Waves In Solids (1st ed.). CRC Press. https://doi.org/10.1201/9780429501012.
%
\bibitem{Gruner_RevModPhys_1988}
G. Grüner, Rev. Mod. Phys. \textbf{60}, 1129 (1988)
%
\bibitem{Knuppel_NatCommun_2025}
P. Knüppel, J. Zhu, Y. Xia, Z. Xia, Z. Han, Y. Zeng, K. Watanabe, T. Taniguchi, J. Shan, and K. F. Mak, Nat. Commun. \textbf{16}, 1959 (2025).
%
\bibitem{Xu_arxiv_2026}
F. Xu, Z. Sun, J. Li, C. Zheng, C. Xu, J. Gao, T. Jia, Y. Su, K. Watanabe, T. Taniguchi, B. Tong, L. Lu, J. Jia, Z. Shi, S. Jiang, J. Lin, Y. Zhang, Y. Zhang, S. Lei, X. Liu, and T. Li, arXiv:2504.06972v2.
%
\bibitem{Wolff_arxiv_2026}
F. K. Wolff, S. Jiao, H. Park, J. Cai, E. Anderson, X. Liu, C. Wang, T. Taniguchi, K. Watanabe, X. Xu, D. Xiao, S. Vaitiekėnas, arXiv:2610.02493.
%
\bibitem{Wu_Lovorn_PRL_2019}
F. Wu, T. Lovorn, E. Tutuc, I. Martin, and A. H. MacDonald, Phys. Rev. Lett. \textbf{122}, 086402 (2019).
%
\bibitem{Cheng_PNAS_2024}
C. Xu, J. Li, Y. Xu, Z. Bi, and Y. Zhang, Proc. Natl. Acad. Sci. U.S.A. \textbf{121} (8) e2316749121 (2024).
%
\bibitem{supple}
The supplementary material contains the sign convention for the phase related to the hopping, details of the $\widetilde{\mathrm{VP_2}}$, $\widetilde{\mathrm{IVC_1}}$, and $\widetilde{\mathrm{IVC_2}}$ phases.
%
\bibitem{BruusFlensberg_2004}
H. Bruus and K. Flensberg, Many-Body Quantum Theory in Condensed Matter Physics: An Introduction, Oxford University Press, Oxford, England, 2004.
%
\bibitem{Misawa_CPC_2019}
T. Misawa, S. Morita, K. Yoshimi, M. Kawamura, Y. Motoyama, K. Ido, T. Ohgoe, M. Imada, and T. Kato, Comput. Phys. Commun. \textbf{235}, 447 (2019).
%
\bibitem{BeccaSorella_2017}
F. Becca and S. Sorella, Quantum Monte Carlo Approaches For Correlated Systems, Cambridge University Press, Cambridge, England, 2017.
%
\bibitem{Toulouse_2016}
J. Toulouse, R. Assaraf, and C. J. Umrigar, in Electron Correlation in Molecules–ab initio Beyond Gaussian Quantum Chemistry, Adv. Quantum Chem., Vol. \textbf{73}, edited by P. E. Hoggan and T. Ozdogan (Academic, New York, 2016), pp. 285–314.
%
\bibitem{Foulkes_RevModPhys_2001}
W. M. C. Foulkes, L. Mitas, R. J. Needs, and G. Rajagopal, Rev. Mod. Phys. \textbf{73}, 33 (2001). 
%
\bibitem{Ceperley_PRB_1977}
D. Ceperley, G. V. Chester, and M. H. Kalos, Phys. Rev. B \textbf{16}, 3081 (1977).
%
\bibitem{Yokoyama_Japan_1987}
H. Yokoyama and H. Shiba, J. Phys. Soc. Jpn. \textbf{56}, 1490-1506 (1987).
%
\bibitem{Biborski_PRB_2020}
A. Biborski, M. Zegrodnik, and J. Spałek, Phys. Rev. B \textbf{101}, 214504 (2020).
%
\bibitem{TaharaImada_2008}
D. Tahara and M. Imada, J. Phys. Soc. Jpn. 77, 114701 (2008).
%
\bibitem{Hotta_PRB_2006}
C. Hotta, N. Furukawa, Phys. Rev. B \textbf{74}, 193107 (2006).
%
\bibitem{Merino_PRL_2013}
J. Merino, A. Ralko, S. Fratini, Phys. Rev. Lett. \textbf{111}, 126403 (2013).

  
\end{thebibliography}
\end{document}